\documentclass[useAMS,usenatbib]{mn2e}
\usepackage{graphicx}
\usepackage{amsmath}
\usepackage{amssymb}

\title[Diffuse Gas in Retired Galaxies]{Diffuse Gas in Retired Galaxies: Nebular Emission Templates and Constraints on the Sources of Ionization}
\author[J. Johansson et al.]{Jonas Johansson$^{1,2}$\thanks{E-mail: jonas.h.johansson@gmail.com}, Tyrone E. Woods$^{1,3}$, Marat Gilfanov$^{1,4,5}$, Marc Sarzi$^{6}$, \newauthor Yan-Mei Chen$^{7,8}$ and Kyuseok Oh$^{9}$\\
$^1$Max-Planck Institut f{\"u}r Astrophysik, Karl-Schwarzschild-Str. 1, D-85741 Garching, Germany\\
$^2$Department of Biological and Environmental Sciences, University of Gothenburg, Medicinaregatan 18, 40530 Gothenburg, Sweden\\
$^3$Monash Centre for Astrophysics, School of Physics and Astronomy, 19 Rainforest Walk, Monash University, VIC, 3800, Australia\\
$^4$Space Research Institute, Profsoyuznaya 84/32, 117997 Moscow, Russia\\
$^5$Kazan Federal University, Kremlevskaya str.18, 420008 Kazan, Russia\\
$^6$Centre for Astrophysics Research, University of Hertfordshire, Hatfield, AL10 9AB, UK\\
$^7$Department of Astronomy, Nanjing University, Nanjing 210093, China\\
$^8$Key Laboratory of Modern Astronomy and Astrophysics (Nanjing University), Ministry of Education, Nanjing 210093, China\\
$^9$Institute for Astronomy, Department of Physics, ETH Z{\"u}rich, Wolfgang-Pauli-Strasse 27, CH-8093, Z{\"u}rich, Switzerland\\
}

\begin{document}

\pagerange{\pageref{firstpage}--\pageref{lastpage}} \pubyear{2013}
\maketitle
\label{firstpage}
\begin{abstract} 
We present emission line templates for passively evolving (``retired") galaxies, useful for investigation of the evolution of the ISM in these galaxies, and characterization of their high-temperature source populations. The templates are based on high signal-to-noise ($>800$) co-added spectra ($3700-6800$\AA) of $\sim11500$ gas-rich Sloan Digital Sky Survey galaxies devoid of star-formation and active galactic nuclei. Stacked spectra are provided for the entire sample and sub-samples binned by mean stellar age. In Johansson~et al (2014), these spectra provided the first measurements of the He II 4686\AA\ line in passively-evolving galaxies, and the observed He II/H$\beta$ ratio constrained the contribution of accreting white dwarfs (the ``single-degenerate'' scenario) to the type Ia supernova rate. In this paper, the full range of unambiguously detected emission lines are presented. Comparison of the observed [O I] 6300\AA/H$\alpha$ ratio with photoionization models further constrains any high-temperature single-degenerate scenario for type Ia supernovae (with 1.5 $\lesssim$ T/$10^{5}K$ $\lesssim$ 10) to $\lesssim$3--6\% of the observed rate in the youngest age bin (i.e. highest SN Ia rate). Hence, for the same temperatures, in the presence of an ambient population of post-AGB stars, we exclude additional high-temperature sources with a combined ionizing luminosity of $\approx 1.35\times 10^{30} L_{\odot}/M_{\odot,*}$ for stellar populations with mean ages of 1 -- 4 Gyrs.  Furthermore, we investigate the extinction affecting both the stellar and nebular continuum. The latter shows about five times higher values. This contradicts isotropically distributed dust and gas that renders similar extinction values for both cases.
\end{abstract}

\begin{keywords}
galaxies: ISM, elliptical and lenticular, cD
\end{keywords}

\section{Introduction}

Historically considered gas-poor, a large fraction of early-type galaxies are now known to exhibit significant fractions of diffuse ionized gas \citep[e.g.][]{caldwell84,phillips86,kim89,goudfrooij94a,sarzi06}. 
However, the sources ionizing the gas present in these objects are still the subject of a long standing debate. In star-burst galaxies, young stars in H~{\sc ii}-regions dominate the ionizing background. 
It is less clear what sources are responsible for ionization of the extended emission-line regions of many early-type, passively-evolving galaxies. Proposed models include heat transfer from hot to cold gas, shock waves, and ionizing radiation from either nuclear activity or a population of evolved stellar sources. The latter option is currently believed to be the best fit thus far on a variety of grounds, with post-asymptotic giant branch (pAGB) stars\footnote{Since pre-planetary (pre-PNe) nebula objects are also commonly referred to as pAGB objects \citep[e.g.][]{vanWinckel03},  
the term `hot evolved low-mass stars' (HOLMES) has lately been suggested for the objects considered in this work \citep[e.g.][]{fernandes11}.} presently the most likely candidate \citep[e.g.][]{sarzi10}.

The energy source(s) powering nebular emission can be traced through the measurement of line ratios sensitive to the hardness of the ionizing continuum. The BPT-diagram \citep*{BPT} is a classic identification tool, based on line ratios between forbidden and recombination lines ([O~{\sc iii}]$\lambda$5007/H$\beta$, [N~{\sc ii}]$\lambda$6584/H$\alpha$, and [S II] $\lambda$6731/H$\alpha$). The BPT-diagram is very effective in separating star-forming galaxies from AGN hosts. In classical BPT-diagrams, it is more difficult to distinguish between Seyfert and LINER hosts, 
but it has been demonstrated that this separation can simply be done using the equivalent width (EW) of H$\alpha$-emission \citep{fernandes11}. Moreover, this method is also very useful in identifying gas ionized by the diffuse background supplied by old passively-evolving stellar populations. 

Further line ratios are useful in assessing the physical conditions within the ionized gas, e.g. the strength of the [S~{\sc ii}]$\lambda$6717/6731 doublet can help separate dense H~{\sc ii} regions from regions of diffuse ionized gas. Hence, a large number of emission lines spanning a wide wavelength range is desired in order to fully constrain the ionizing background. For the study of early-type galaxies containing significantly less gas than star-forming galaxies, the quality of the data is particularly important for moderate to weak emission lines to be detected. The final aspect that must be considered is the sample selection. In order to  study the gas in retired galaxies, it is essential to exclude as carefully as possible any star forming or AGN galaxies from our sample.

In \citet{johansson14} we co-added thousands of SDSS spectra from a carefully selected sample of galaxies devoid of star formation and AGNs, in order to detect the weak He~{\sc ii}~$\lambda$4686 line as a tracer of accreting WD SN~Ia progenitors. We found that the observed strength of this line is consistent with pAGBs being the sole ionizing source, and indicates that low-temperature accreting, steadily nuclear-burning white dwarfs can account for a maximum of $\sim$10 per cent of the SN~Ia rate. In the process of fitting the HeII$\lambda$4686 line we simultaneously fit a wide range of emission lines from the far UV to the NIR. Thanks to the very high quality of the stacked spectra, the range of detected emission lines provides excellent templates for the emission lines of retired galaxies to further constrain the sources of ionizing radiation in these stellar populations. 

The nature of dust in passively evolving galaxies also remains poorly understood. While dust has been well probed in star-forming galaxies, the same cannot be said in passively evolving galaxies. In the process of fitting emission lines, we simultaneously fit templates to the stellar continuum. This gives us the opportunity to investigate the extent of dust reddening affecting the stellar continuum and nebular emission separately. In star-forming galaxies, the dust reddening affecting the nebular emission has been found to typically be twice as high as that affecting the stellar continuum, suggesting that young stars in dust enshrouded H~{\sc ii}-regions dominate the nebular ionization. In a diffuse medium the gas is believed to be more isotropically distributed, resulting in a similar dust reddening of the nebular and stellar emission. Controversially, for the retired galaxies studied in this work, we find that the relative difference between dust reddening affecting the nebular and stellar emission is even greater than that found in star-forming galaxies.

This paper is organized as follows. In Section~2 the data and fitting procedure are introduced. We present the emission line templates and dust reddening of retired galaxies in Section~3. The emission line templates are compared to photoionisation models in Section~4 together with a discussion on dust reddening in retired galaxies. Concluding remarks are given in Section~5.

\section{Data sample, Co-addition and emission line fitting}
\label{datasection}

Summaries of the sample selection (Section~\ref{data}) and stacking technique (Section~\ref{stacking}) are given here and the reader is referred to \citet{johansson14} for a detailed description. In general, the emission line fitting follows \citet{johansson14} and is summarized in Section~\ref{stacking:EMfitting}, but we have extended the wavelength range used and added additional fitting to estimate the amount of extinction. These extensions are described in detail in Section~\ref{stacking:EMfitting}.

\subsection{Data sample}
\label{data}

The sample used in this study is selected from the SDSS-II data release 7 \citep[DR7][]{DR7}, which
include spectroscopy for more than 930~000 galaxies. A number of selection criteria  
are applied to the full DR7 sample 

\begin{figure}
\centering
\includegraphics[height=0.35\textheight,clip=true,trim=0cm 3cm 4cm 2cm,angle=180]
{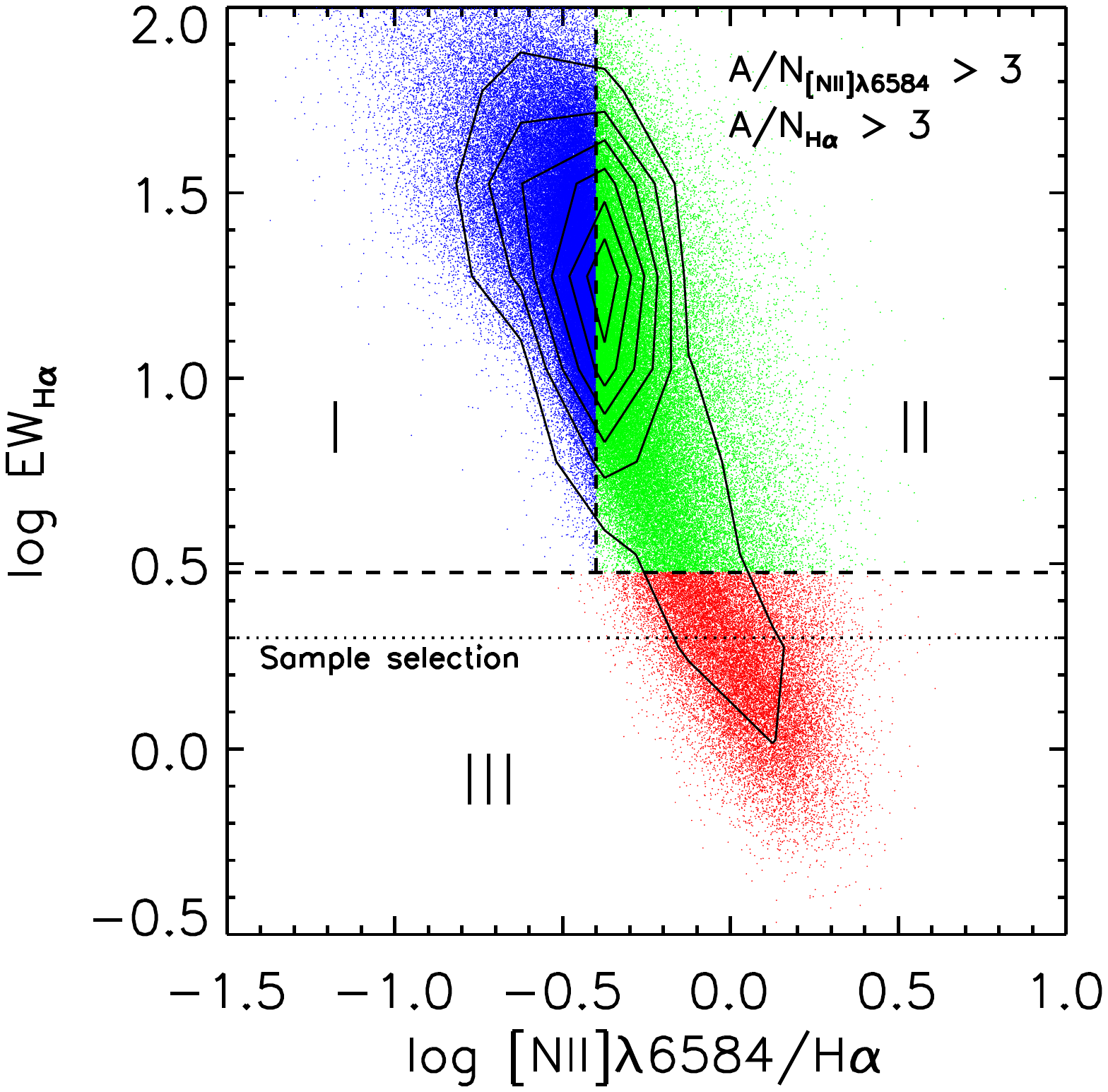}\\
\caption{The WHAN-diagram \citep{fernandes11}, [N~{\sc ii}~]$\lambda$6584/H$\alpha$ line ratio versus the
  strength of H$\alpha$ (quantified by its equivalent width), for SDSS
  galaxies with detected H$\alpha$ and [N~{\sc ii}~]$\lambda$6584 emission, i.e. for
  which the A/N~$>3$ for both lines. The regions defined by the dashed lines
  include galaxies with ionizing UV-radiation dominated by young stars
  in star-forming (SF) regions (blue data points in the top left region denoted I), an AGN (green data points in the top right region denoted II) or UV-bright
  sources in old stellar populations (red data points in the bottom region denoted III). The dotted line
  represents our selection of old, so-called retired galaxies.}
\label{selection}
\end{figure}

\begin{itemize}
\item[1.] \textit{Redshift cut}, selecting galaxies with redshifts in the range  $0.04<z<0.1$, leaving 305~598 objects
\item[2.] \textit{WHAN} \citep{fernandes11} \textit{cut}, selecting gas rich galaxies where the ionizing photon field is not powered by young OB-stars or accreting black holes by using the following criteria for the strengths of emission lines measured in \citet{oh11}; $\log\textrm{EW}_{\textrm{H}\alpha}<0.3$ and H$\alpha$ and [N~{\sc ii}~]$\lambda$6584 must be detected with an amplitude-over-noise ratio (A/N)$>3$. These cuts are illustrated in Fig.~\ref{selection} (a remake of fig.~1 in \citet{johansson14}) and leave a sample of 13 875 galaxies. 
\item[3a.] \textit{PCA cut}, selecting galaxies without star formation in the last Gyr, by applying the principal component analysis (PCA) technique of \citet{chen12} which exploits spectral features sensitive to the fraction of stars formed in the last Gyr. Objects where more than one per cent of the stellar mass has formed in a recent star formation episode in the last Gyr are excluded, resulting in a final sample of 11~593 galaxies.
\item[3b.] \textit{NUV-$r$ cut}, complementary to the PCA technique, we separately exclude galaxies with recent star formation by combining  near ultra-violet (NUV) photometry from the galaxy evolution explorer (GALEX) database \citep{martin05} with SDSS $r$ band magnitudes. Objects with any star formation in the last Gyr typically have NUV-$r$ colours $<5.5$ \citep{kaviraj07}. This selection criterion is more conservative, and should be considered a complement to confirm the results for the PCA selected sample,  leaving a final sample of 4~061 galaxies.
\end{itemize}

The final galaxy samples \textit{PCA cut} and \textit{NUV-$r$ cut} are divided into several bins of average stellar population age. The age is derived for each individual spectrum using the technique based on absorption line indices from \citet{johansson12}.  Only a brief description of this method is given here, while the reader is referred to \citet{johansson12} and  \citet{johansson14} for details.
Observed Lick indices, which are defined for 25 prominent absorption features in the optical \citep{worthey94}, are taken from the OSSY database \citep{oh11}. These are compared with the modeled absorption line indices from \citet*{TMJ11} through a $\chi^2$-minimization routine to find a best fitting set of parameters including stellar population age, metallicity and element abundance ratios.

The samples are divided in age bins containing at least $\sim2000$ objects,  which is enough to reach the detection limit of the  He~{\sc ii}~$\lambda$4686 lines at a S/N of $\sim500$ of the stacked spectra according to \citet{woods}. The age distributions are shown in fig.~3 in \citet{johansson14} {where it was also demonstrated that Lick index derived ages of the stacked spectra agree very well with the chosen age bins}.
The \textit{PCA cut} sample  is divided into four bins ($<4$ Gyr, $4-6$ Gyr, $6-9$ Gyr and $>9$ Gyr). These numbers together with the resulting S/N of the stacked spectra are presented in Table~\ref{table:stacks} \citep[a remake of table~2 in][see details therein]{johansson14}. The \textit{NUV-$r$ cut} sample is divided into two age bins ($<6$ Gyr and $>6$ Gyr), leaving $\sim2000$ galaxies in each bin. {For each age bin, we now also provide the summed MPA-JHU fibre masses from the SDSS DR7\footnote{Derived from photometry following \cite{kauffmann03,salim07}, see http://www.mpa-garching.mpg.de/SDSS/DR7/} and the sum of each fibre mass over luminosity distance squared (for a WMAP7 cosmology). These values allow one to make a direct comparison between spectral synthesis models and our results for the observed line fluxes in the following section.}

\begin{table}
\center
\caption{ Salient features of each of the co-added spectra analysed in
  this work. The columns give the age range of the galaxies in each group (col.~1), method for discarding galaxies with stars formed in the last Gyr (col.~2), number
  of spectra in each group (col.~3) and resulting S/N for the stacked
  spectra (col.~4) computed using the propagation of the
formal SDSS uncertainties (see text). {We also include the summed masses (MPA-JHU) and the summed mass over luminosity distance squared (for a WMAP7 cosmology) for each stack.}}
  \label{table:stacks}
\begin{tabular}{ccrrrr}
\hline
 \multicolumn{2}{c}{\bf Stack }  & \bf N~~ & \bf S/N & $\Sigma (\rm{M}_{*}$) & $\Sigma (\rm{M}_{*}/\rm{D}_{\rm{L}}^{2})$\\
& & & & $\rm{M}_{\odot}$ & $\rm{M}_{\odot}/\rm{pc}^{2}$\\
 \hline
All ages  & PCA &11593 & 1751 & 2.2e+14 & 2.5e-3\\
$<$4 Gyr & PCA & 2740  &  882 & 4.6e+13 & 4.8e-4\\ 
4-6  Gyr & PCA & 3380  & 1044 & 6.2e+13 & 6.8e-4\\
6-9  Gyr & PCA & 2729  & 1000 & 5.5e+13 & 6.4e-4\\
$>$9 Gyr & PCA & 2744  & 1018 & 6.0e+13 & 7.0e-4\\ 
$<$6 Gyr & NUV-$r$ & 1953 & 829 & 3.7e+13 & 4.2e-4\\
$>$6 Gyr & NUV-$r$ & 2108 & 951 & 4.9e+13 & 6.1e-4\\
\hline
\end{tabular}
\end{table}
\subsection{Co-addition of individual Spectra}
\label{stacking}

We apply a number of corrections to the individual spectra before stacking them. To correct for small-scale fluctuations, the flux calibration of \citet{yan11} is applied. The spectra are then de-reddened for Galactic extinction using the extinction curve of diffuse gas from \citet{odonnell94}, by applying an R$_v$ value of 3.1 and the Galactic E(B--V) values from the extinction maps of \citet*{schlegel98}. Finally, the spectra are brought to rest-frame wavelengths and to a common wavelength grid.

For each sample selection, we then co-add the spectra in emitted flux and the corresponding statistical error arrays are co-added in quadrature. The mean S/N around 5000 \AA\ ($\pm250$ \AA) for all sub-samples are presented in Table~\ref{table:stacks} (col.~4)  and vary between $\sim830$ and $\sim1750$. As noted in \citet{johansson14}, these values should be considered approximations for characterizing the quality of the stacked spectra. 

\subsection{Emission line fitting and dust extinction}
\label{stacking:EMfitting}

The fitting code {\sc gandalf} \citep{sarzi06} is adopted for measuring emission lines. The reader is referred to \citet{sarzi06} for a detailed description, while only a brief description of the code is given here. {\sc gandalf} is very efficient in the separation of emission and absorption spectra of composite galaxy spectra. This task is mainly divided into two steps, where the considered emission lines are first masked while the {\sc ppxf}
code \citep{ppxf} measures kinematic broadening of the absorption spectrum by fitting a set of stellar templates. In the second step, the emission line mask is lifted and the stellar kinematics are fixed. {\sc gandalf} then refits the stellar continuum through the simultaneous fitting of Gaussian emission line templates of both re-combination and forbidden lines.

The full medium-resolution
Isaac Newton telescope library \citep[MILES,][]{miles}, which includes observed spectra for 985 solar neighborhood stars, is adopted as stellar templates. This stellar library is chosen as it covers a particularly large stellar parameter space, such that a very accurate fit of the stellar continuum can be achieved \citep{capp07,sarzi10}. 
Since the MILES library has a fixed instrumental full width half maximum (FWHM) resolution of $\Delta\lambda\sim2.5$ \AA\ \citep{beifiori11} and the SDSS resolution is fixed at R~$\sim2000$, both the SDSS and MILES spectra are degraded to match each other prior to running {\sc gandalf}, 

For the initial goal of this work to detect the He\,{\sc ii}\,$\lambda$4686 line, we used the wavelength window 3800--6800 \AA. We have now expanded this window to also cover the  [O\,{\sc ii}]$\lambda$3727 doublet. Hence, the range 3700--6800 \AA\ is instead adopted and we have re-run {\sc gandalf} for all stacked spectra. We confirm that the previous results have not changed and that the emission line fluxes vary with $<1\sigma$ of the flux errors when adopting the two different wavelength windows. The chosen wavelength range includes a range of strong absorption
features that can constrain the best-fitting set of stellar
templates. The adopted wavelength window also covers a wide range of emission lines including the strong
H$\alpha$ and
[N\,{\sc ii}\,]$\lambda$6584 lines that can be used for constraining the position and
Gaussian width of weaker emission lines, e.g., He\,{\sc ii}\,$\lambda$4686. Hence, the kinematics of all forbidden
lines, including the [O\,{\sc iii}\,]$\lambda$4959,$\lambda$5007 doublet, are tied to the kinematics of 
[N\,{\sc ii}\,]$\lambda$6584, while all recombination lines are
tied to that of H$\alpha$.

\begin{figure}
\centering
\includegraphics[height=0.27\textheight,clip=true,trim=0cm 2cm 0cm 1cm,angle=180]{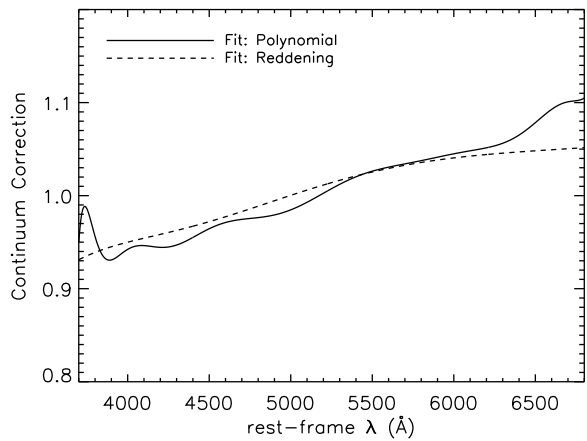}\\
\caption{Comparison of the two means of continuum correction applied in this work, as a function of wavelength over the spectral window of interest. The solid line denotes the cumulative correction found using a 15 degree multiplicative polynomial, while the dashed line denotes the same found using the extinction curve of O'Donnell (1994).}
\label{fig:contcorr}
\end{figure}

One freedom in the setup of {\sc gandalf} is the adjustment of the continuum of the stellar templates. As the best-fitting stellar template is constrained using strong absorption features the stellar continuum is simultaneously adjusted to fit the overall shape of the observed spectrum. 
{\sc gandalf} allows either a reddening law or multiplicative polynomials to be used. Concerning the latter option, an arbitrary multiplicative degree can be used and a higher precision of the continuum can be achieved compared to using a reddening law. 
With the main aim of detecting the weak He\,{\sc ii}\,$\lambda$4686 line in \citet{johansson14}, we increased the multiplicative degree until the $\chi^2$ of the fit around the position of He\,{\sc ii}\,$\lambda$4686 ($\pm$200 \AA) was no longer reduced. This occurred for a degree of 15 \citep[see fig.~4 in][]{johansson14}, which affects the continuum shape at a 200~\AA\ level for the adopted wavelength range. To test the robustness of the derived emission lines and to estimate the amount of extinction affecting the stellar continuum, we now also run {\sc gandalf} using reddening to adjust the flux of the stellar templates through
\begin{equation}
\textrm{F}_\textrm{obs}(\lambda)=\textrm{F}_\textrm{int}(\lambda)\times10^{-0.4\textrm{R}_v\textrm{E}(B-V)k(\lambda)}
\label{eq:adjust}
\end{equation}
where F$_\textrm{obs}$ and F$_\textrm{int}$ are the adjusted (or "observed") and intrinsic fluxes, respectively, and $k$ is the extinction curve. We adopt the extinction curve from \citet{odonnell94} derived for diffuse gas in the Milky Way and an R$_v=3.1$. The color excess E($B-V$) is a free parameter in the fitting.

For all the fitted emission lines we determine the level of detection using the amplitude-over-noise (A/N) ratio, i.e., 
between the peak amplitude of the modeled Gaussian and the residual of the fit in a wavelength window of $\pm$200~\AA\ around the considered line \citep[for more details see][]{johansson14}. 
To produce a reliable template of emission lines for retired galaxies, we only consider lines that are unambiguously detected and, thus, adopt a conservative detection threshold of A/N~$>6$. 
By scaling the statistical errors to get a $\chi^2\sim1$ for the overall fit \citep{sarzi05,johansson14}, the estimated errors on the flux of the fitted emission lines account for contributions from both statistical and systematic errors.

Using the observed Balmer decrements we can estimate the amount of intrinsic reddening of the gas for our stacked spectra, which can be compared to the extinction affecting the stellar continuum (see Section~\ref{results:dust}). 
From the observed Balmer line ratio R~$=$~H$\alpha$/H$\beta$ we derive the dust reddening 
affecting the nebular emission using the standard definition of the color excess \citep[e.g.][]{calzetti97}
\begin{equation}
\textrm{E}(B-V)_g=\frac{\log_{10}(\textrm{R}_\textrm{obs}/\textrm{R}_\textrm{int})}{0.4\textrm{R}_v[k(\lambda_{\textrm{H}\beta})-k(\lambda_{\textrm{H}\alpha})]}
\label{eq:ebmv}
\end{equation}
where R$_\textrm{obs}$ and R$_\textrm{int}$ are the observed and intrinsic Balmer line ratios, respectively, and $k(\lambda)$ is the extinction curve.  For the latter we once again adopt the extinction curve for diffuse gas from \citet{odonnell94}. As default we adopt R$_v=3.1$, typically found in the Milky Way, and an atomic Balmer line ratio R$_\textrm{int}=2.87$ for a typical gas temperature of 10~000 K \citep{osterbrock89}. 

\begin{figure*}
\centering
\includegraphics[height=0.55\textheight,clip=true,trim=1cm 2cm 2.5cm 2cm,angle=180]{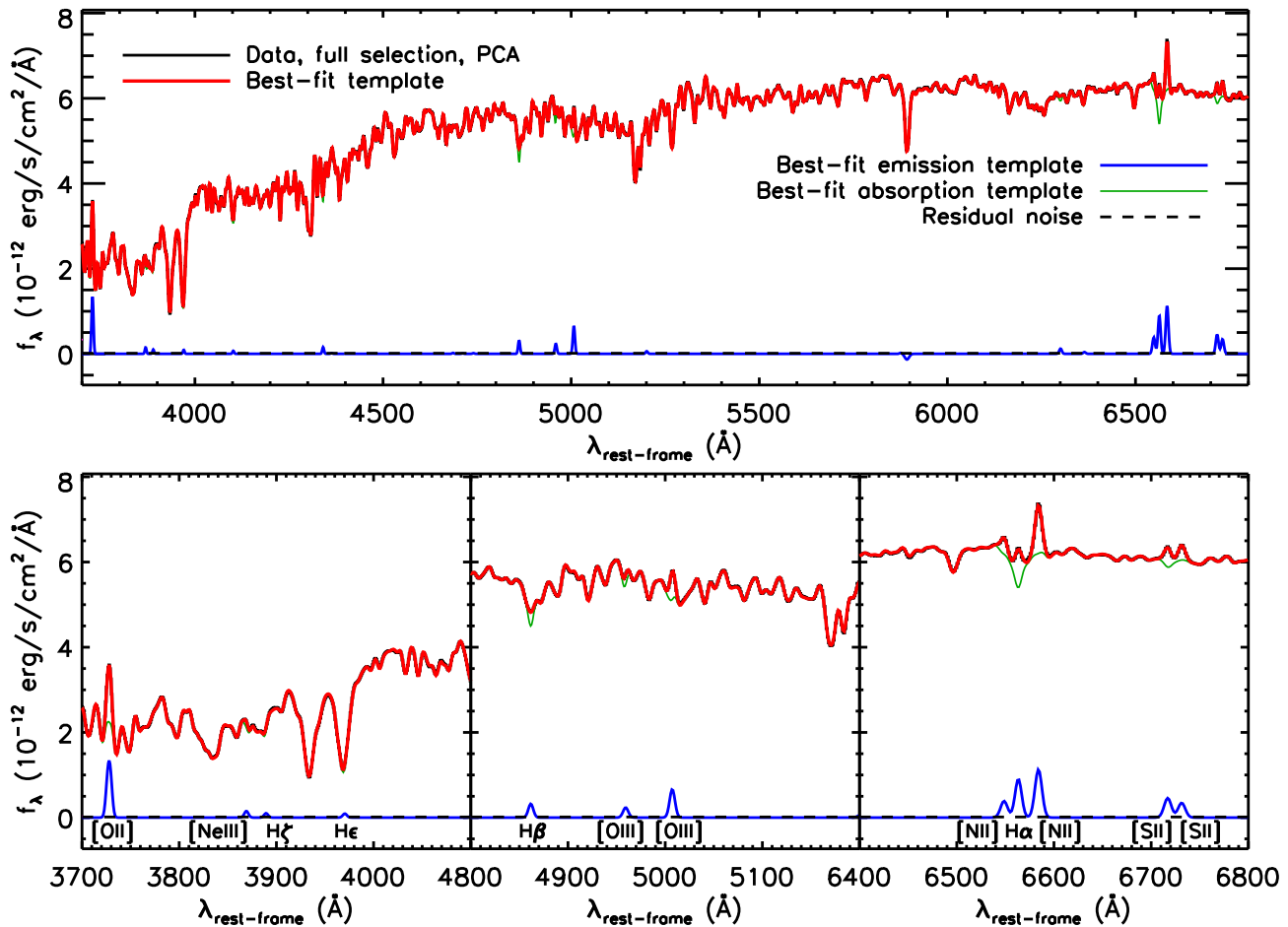}\\
\caption{{Data and model of stacked spectrum. Upper panel: total spectrum obtained by co-adding the spectra
  of all galaxies in our sample of quiescent objects (11593
  galaxies) together with our best-fitting model. The
  stellar component of this model is shown by the red line,
  whereas our best-fitting nebular spectrum is shown in blue. Lower
  panels: zoom-in illustrating the difference between the stacked spectrum and best-fitting
  stellar model around the position of key diagnostic lines,
  together with our best-fitting emission-line model (blue line). The position of all lines is indicated with the appropriate labels.}}
\label{fit:HaLT0p4}
\end{figure*}

{In fact, when using a screen-like reddening to adjust for the stellar continuum, {\sc gandalf} also uses an additional `nebular' reddening component to reproduce the observed H$\alpha$/H$\beta$ ratio starting from a preset intrinsic ratio between all recombination lines, so that an estimate for E($B-V$)$_g$ is directly returned in this case \citep[see also][for details]{oh11}. On the other hand, when adopting a polynomial correction for the stellar continuum, the strength of the Balmer lines are let free to adjust themselves until they match the data, so in this case we estimate E($B-V$)$_g$ from Eq~\ref{eq:ebmv}. Clearly, in either case both H$\alpha$ and H$\beta$ have to be well detected in order not to incur in biased and unphysical line ratios and reddening values, which is the case for all our stacked spectra.}

\begin{table*}
\center
\caption{{Observed and reddening-corrected total emission line fluxes for unambiguously detected lines in each of our stacked spectra. Errors for both the observed ($\sigma_{obs}$) and reddening-corrected ($\sigma_{corr}$) fluxes are reported.}}
\label{emisson}
\begin{tabular}{rlcrrrrrrr}
\hline
\bf N  &  \bf Line & \bf $\lambda$ (\AA) &              \multicolumn{1}{c}{\bf  $<$ 4 Gyr}                     &                    \multicolumn{1}{c}{\bf 4-6 Gyr}                      &                   \multicolumn{1}{c}{\bf 6-9 Gyr}                      &                  \multicolumn{1}{c}{\bf  $>$ 9 Gyr}                  &                     \multicolumn{1}{c}{\bf $<$ 6 Gyr}             &                        \multicolumn{1}{c}{\bf $>$ 6 Gyr}                   &                    \multicolumn{1}{c}{\bf Full sel.}  \\      
     &           &                              &				\multicolumn{1}{c}{(PCA)}			&			\multicolumn{1}{c}{(PCA)}			&			\multicolumn{1}{c}{(PCA)}				&			\multicolumn{1}{c}{(PCA)}				&			\multicolumn{1}{c}{(NUV-r)}				&				\multicolumn{1}{c}{(NUV-r)} 			&						\\
     &           &                              &     \multicolumn{1}{c}{obs.$^a$}        &        \multicolumn{1}{c}{obs.$^a$}   &          \multicolumn{1}{c}{obs.$^a$}     &       \multicolumn{1}{c}{obs.$^a$}      &      \multicolumn{1}{c}{obs.$^a$}      &     \multicolumn{1}{c}{obs.$^a$}          &   \multicolumn{1}{c}{obs.$^a$}   \\    
     &           &                              &     \multicolumn{1}{c}{corr.$^b$}        &        \multicolumn{1}{c}{corr.$^b$}   &          \multicolumn{1}{c}{corr.$^b$}     &        \multicolumn{1}{c}{corr.$^b$}      &      \multicolumn{1}{c}{corr.$^b$}     &     \multicolumn{1}{c}{corr.$^b$}          &   \multicolumn{1}{c}{corr.$^b$}   \\ 
     &           &                              &     \multicolumn{1}{c}{$\sigma_{obs}/\sigma_{corr}$}        &        \multicolumn{1}{c}{$\sigma_{obs}/\sigma_{corr}$}   &          \multicolumn{1}{c}{$\sigma_{obs}/\sigma_{corr}$}     &        \multicolumn{1}{c}{$\sigma_{obs}/\sigma_{corr}$}      &      \multicolumn{1}{c}{$\sigma_{obs}/\sigma_{corr}$}      &     \multicolumn{1}{c}{$\sigma_{obs}/\sigma_{corr}$}          &   \multicolumn{1}{c}{$\sigma_{obs}/\sigma_{corr}$}   \\
\hline
   1  &    [O\,{\sc ii}]  &   3726.03    &  66.1 &     94.1 &     88.4 &    109.4 &     58.4 &     90.6 &    404.3 \\
        &        &                   &     213.4 &    305.4 &    259.5 &    363.7 &    222.8 &    283.4 &   1278.4 \\
        &    &                   &      $\pm$1.7/$\pm$18.3 & $\pm$2.3/$\pm$28.9 & $\pm$2.4/$\pm$28.9 & $\pm$3.4/$\pm$46.9 & $\pm$1.7/$\pm$21.9 & $\pm$2.7/$\pm$36.6 & $\pm$12.6/$\pm$122.3 \\
   2  &    [O\,{\sc ii}]  &   3728.73     & 71.0 &     98.7 &     95.9 &    119.8 &     65.1 &     96.6 &    448.8\\
        &           &                   &  229.3 &    320.1 &    281.4 &    397.9 &    248.1 &    302.0 &   1418.2\\
        &        &                   &  $\pm$1.7/$\pm$19.6 & $\pm$2.3/$\pm$30.2 & $\pm$2.4/$\pm$31.3 & $\pm$3.3/$\pm$51.3 & $\pm$1.5/$\pm$24.4 & $\pm$2.6/$\pm$39.0 & $\pm$11.1/$\pm$135.6\\
   3  &    [Ne\,{\sc iii}]  & 3868.69   &   14.3 &     18.9 &     18.0 &     21.7 &     12.5 &     18.2 &     80.6\\
        &                &               &  44.6 &     59.2 &     51.2 &     69.7 &     45.9 &     55.0 &    246.0\\
        &                &       &      $\pm$1.0/$\pm$3.7 & $\pm$1.3/$\pm$5.4 & $\pm$1.3/$\pm$5.5 & $\pm$1.6/$\pm$8.7 & $\pm$0.8/$\pm$4.4 & $\pm$1.3/$\pm$6.9 & $\pm$5.8/$\pm$22.8\\
   4   &   H$\gamma$    &     4340.46   &   16.0 &     21.5 &     20.4 &     21.8 &     12.4 &     18.6 &     89.4\\
       &                &               &   45.3 &     61.2 &     53.2 &     63.2 &     40.7 &     51.3 &    248.4\\
       &                &             & $\pm$0.8/$\pm$3.4 & $\pm$1.1/$\pm$5.1 & $\pm$1.1/$\pm$5.2 & $\pm$1.4/$\pm$7.2 & $\pm$0.7/$\pm$3.5 & $\pm$1.2/$\pm$5.9 & $\pm$4.7/$\pm$20.9\\
   5    &  H$\beta$   &       4861.32 &     37.7 &     47.9 &     43.0 &     48.2 &     28.8 &     40.5 &    205.1\\
       &                &               &   94.4 &    120.6 &    100.1 &    123.6 &     82.1 &     99.1 &    505.8\\
       &                &            &  $\pm$0.8/$\pm$6.3 & $\pm$1.1/$\pm$8.8 & $\pm$1.2/$\pm$8.7 & $\pm$1.5/$\pm$12.4 & $\pm$0.7/$\pm$6.3 & $\pm$1.3/$\pm$10.0 & $\pm$4.9/$\pm$37.6\\
   6    &  [O\,{\sc iii}] &   5006.77 &     74.1 &    106.2 &     99.9 &    114.1 &     65.9 &     94.7 &    455.9\\
       &                &               &  179.3 &    257.8 &    225.1 &    282.1 &    180.7 &    223.7 &   1085.6\\
       &                &       &       $\pm$0.9/$\pm$11.6 & $\pm$1.2/$\pm$18.4 & $\pm$1.3/$\pm$18.9 & $\pm$1.6/$\pm$27.4 & $\pm$0.8/$\pm$13.4 & $\pm$1.3/$\pm$21.8 & $\pm$5.3/$\pm$78.2\\
   7    &  [O\,{\sc i}]  &    6300.20 &     17.4 &     23.8 &     23.5 &     28.2 &     15.8 &     23.4 &    105.4\\
     &                &               &     33.7 &     46.2 &     43.0 &     55.4 &     33.6 &     44.5 &    201.6\\
     &                &               & $\pm$0.8/$\pm$1.6 & $\pm$1.2/$\pm$2.5 & $\pm$1.2/$\pm$2.7 & $\pm$1.6/$\pm$4.0 & $\pm$0.7/$\pm$1.9 & $\pm$1.3/$\pm$3.2 & $\pm$5.1/$\pm$10.9\\
   8    &  H$\alpha$   &      6562.80 &    143.6 &    182.9 &    160.2 &    185.1 &    114.1 &    153.3 &    777.8\\
     &                &               &    270.1 &    345.0 &    286.3 &    353.6 &    234.9 &    283.5 &   1446.6\\
     &                &               & $\pm$1.0/$\pm$12.3 & $\pm$1.4/$\pm$17.4 & $\pm$1.5/$\pm$17.1 & $\pm$2.0/$\pm$24.4 & $\pm$0.9/$\pm$12.4 & $\pm$1.7/$\pm$19.6 & $\pm$6.2/$\pm$73.9\\
  9    &   [N\,{\sc ii}]  &   6583.34 &    178.1 &    237.6 &    219.1 &    253.6 &    153.5 &    211.7 &   1022.5\\
     &                &               &    334.3 &    447.2 &    390.8 &    483.5 &    315.0 &    390.6 &   1897.7\\
     &                &               & $\pm$1.1/$\pm$15.4 & $\pm$1.5/$\pm$22.7 & $\pm$1.6/$\pm$23.4 & $\pm$2.0/$\pm$33.5 & $\pm$0.9/$\pm$16.6 & $\pm$1.7/$\pm$27.1 & $\pm$6.4/$\pm$97.4\\
  10   &   [S \,{\sc ii}] &   6716.31 &     71.3 &     96.4 &     89.4 &    108.7 &     61.8 &     88.6 &    421.6\\
     &                &               &    132.0 &    179.0 &    157.4 &    204.2 &    124.8 &    161.3 &    771.9\\
     &                &               & $\pm$1.0/$\pm$5.9 & $\pm$1.4/$\pm$8.9 & $\pm$1.5/$\pm$9.2 & $\pm$2.0/$\pm$13.8 & $\pm$0.9/$\pm$6.4 & $\pm$1.7/$\pm$10.9 & $\pm$6.3/$\pm$38.7\\
  11   &   [S\,{\sc ii}] &    6730.68 &     52.9 &     72.2 &     67.5 &     79.8 &     46.1 &     65.7 &    317.9\\
     &                &               &     97.8 &    133.7 &    118.8 &    149.8 &     93.0 &    119.3 &    581.1\\
     &                &               &  $\pm$1.1/$\pm$4.4 & $\pm$1.5/$\pm$6.6 & $\pm$1.5/$\pm$6.9 & $\pm$2.0/$\pm$10.1 & $\pm$0.9/$\pm$4.8 & $\pm$1.7/$\pm$8.1 & $\pm$6.4/$\pm$29.1\\
\hline
\end{tabular}\\
\flushleft
~~~~~~~~~$^a$Observed flux, unit 10$^{-14}$ erg s$^{-1}$ cm$^{-2}$.\\
~~~~~~~~~$^b$Reddening corrected flux, unit 10$^{-14}$ erg s$^{-1}$ cm$^{-2}$.\\
\end{table*}

\begin{table*}
\center
\caption{{Observed and reddening-corrected line intensities relative to H$\beta$. Errors for both the observed ($\sigma_{obs}$) and reddening-corrected ($\sigma_{corr}$) ratios are reported.}}
\label{HbRatio}
\begin{tabular}{lrrrrrrr}
\hline
   \bf Line/H$\beta$ &              \multicolumn{1}{c}{\bf  $<$ 4 Gyr}                     &                    \multicolumn{1}{c}{\bf 4-6 Gyr}                      &                   \multicolumn{1}{c}{\bf 6-9 Gyr}                      &                  \multicolumn{1}{c}{\bf  $>$ 9 Gyr}                  &                     \multicolumn{1}{c}{\bf $<$ 6 Gyr}             &                        \multicolumn{1}{c}{\bf $>$ 6 Gyr}                   &                    \multicolumn{1}{c}{\bf Full sel.}  \\      
                                              &				\multicolumn{1}{c}{(PCA)}			&			\multicolumn{1}{c}{(PCA)}			&			\multicolumn{1}{c}{(PCA)}				&			\multicolumn{1}{c}{(PCA)}				&			\multicolumn{1}{c}{(NUV-r)}				&				\multicolumn{1}{c}{(NUV-r)} 			&						\\
                                              &     \multicolumn{1}{c}{obs.}        &        \multicolumn{1}{c}{obs.}   &          \multicolumn{1}{c}{obs.}     &       \multicolumn{1}{c}{obs.}      &      \multicolumn{1}{c}{obs.}      &     \multicolumn{1}{c}{obs.}          &   \multicolumn{1}{c}{obs.}   \\    
                                              &     \multicolumn{1}{c}{corr.}        &        \multicolumn{1}{c}{corr.}   &          \multicolumn{1}{c}{corr.}     &        \multicolumn{1}{c}{corr.}      &      \multicolumn{1}{c}{corr.}     &     \multicolumn{1}{c}{corr.}          &   \multicolumn{1}{c}{corr.}   \\ 
                                              &     \multicolumn{1}{c}{$\sigma_{obs}/\sigma_{corr}$}        &        \multicolumn{1}{c}{$\sigma_{obs}/\sigma_{corr}$}   &          \multicolumn{1}{c}{$\sigma_{obs}/\sigma_{corr}$}     &        \multicolumn{1}{c}{$\sigma_{obs}/\sigma_{corr}$}      &      \multicolumn{1}{c}{$\sigma_{obs}/\sigma_{corr}$}      &     \multicolumn{1}{c}{$\sigma_{obs}/\sigma_{corr}$}          &   \multicolumn{1}{c}{$\sigma_{obs}/\sigma_{corr}$}   \\
\hline
       {[O\,{\sc ii}]$\lambda$3726/H$\beta$}    &  1.75 & 1.96    & 2.06     & 2.27   & 2.03   & 2.24   & 1.97   \\
                                   &  2.26 &  2.53   & 2.59     & 2.94   & 2.71   & 2.86   & 2.53      \\
                               &      $\pm$0.06/$\pm$0.25 & $\pm$0.07/$\pm$0.30 & $\pm$0.08/$\pm$0.37 & $\pm$0.10/$\pm$0.48 & $\pm$0.08/$\pm$0.34 & $\pm$0.10/$\pm$0.47 & $\pm$0.08/$\pm$0.31 \\
        {[O\,{\sc ii}]$\lambda$3729/H$\beta$}  & 1.88  &  2.06   &  2.23    &  2.49  &  2.26  & 2.39   & 2.19   \\
                                      &  2.43 &  2.66   & 2.81     & 3.22   & 3.02   & 3.05   & 2.80  \\
                                   & $\pm$0.06/$\pm$0.26 & $\pm$0.07/$\pm$0.32 & $\pm$0.08/$\pm$0.40 & $\pm$0.10/$\pm$0.53 & $\pm$0.08/$\pm$0.38 & $\pm$0.10/$\pm$0.50 & $\pm$0.08/$\pm$0.34 \\  
        {[Ne\,{\sc iii}]$\lambda$3869/H$\beta$}   &  0.38  &  0.39   & 0.42     & 0.45   & 0.43   &  0.45  & 0.44  \\
                                       &  0.47 & 0.49    & 0.51     & 0.56   & 0.56   &  0.55  & 0.49  \\
                               &  $\pm$0.03/$\pm$0.05 & $\pm$0.03/$\pm$0.06 & $\pm$0.03/$\pm$0.07 & $\pm$0.04/$\pm$0.09 & $\pm$0.03/$\pm$0.07 & $\pm$0.04/$\pm$0.09 & $\pm$0.03/$\pm$0.06 \\     
        {H$\gamma$/H$\beta$}   & 0.42  &   0.45  &  0.47    & 0.45   &  0.43  & 0.46   & 0.44   \\
                                      &  0.48 &  0.51   & 0.53     &  0.51  &  0.50  &  0.52  &  0.49  \\
                                    &  $\pm$0.02/$\pm$0.05 & $\pm$0.03/$\pm$0.06 & $\pm$0.03/$\pm$0.07 & $\pm$0.03/$\pm$0.08 & $\pm$0.03/$\pm$0.06 & $\pm$0.03/$\pm$0.08 & $\pm$0.03/$\pm$0.06 \\ 
        {[O\,{\sc iii}]$\lambda$5007/H$\beta$} & 1.97  & 2.22    & 2.32     & 2.37   & 2.29   & 2.34   & 2.22     \\
                                      &  1.90 & 2.14    &  2.25    & 2.28   & 2.20   & 2.26   & 2.15  \\
                              &   $\pm$0.05/$\pm$0.18 & $\pm$0.06/$\pm$0.22 & $\pm$0.07/$\pm$0.27 & $\pm$0.08/$\pm$0.32 & $\pm$0.06/$\pm$0.23 & $\pm$0.08/$\pm$0.32 & $\pm$0.06/$\pm$0.22 \\      
        {[O\,{\sc i}]$\lambda$6300/H$\beta$} & 0.46  & 0.50    & 0.55     & 0.59   & 0.55   & 0.58   & 0.51    \\
                                   & 0.36  & 0.38    &  0.43    & 0.45   &  0.41  & 0.45   & 0.40     \\
                                    &  $\pm$0.02/$\pm$0.03 & $\pm$0.03/$\pm$0.03 & $\pm$0.03/$\pm$0.05 & $\pm$0.04/$\pm$0.06 & $\pm$0.03/$\pm$0.04 & $\pm$0.04/$\pm$0.06 & $\pm$0.03/$\pm$0.06 \\
        {H$\alpha$/H$\beta$} & 3.81  & 3.82    &  3.72    & 3.84   & 3.97   & 3.78   & 3.79    \\
                                    &  2.86 & 2.86    &  2.86    & 2.86   & 2.86   & 2.86   & 2.86  \\
                                    &  $\pm$0.09/$\pm$0.23 & $\pm$0.010/$\pm$0.25 & $\pm$0.11/$\pm$0.30 & $\pm$0.13/$\pm$0.35 & $\pm$0.10/$\pm$0.27 & $\pm$0.13/$\pm$0.35 & $\pm$0.10/$\pm$0.26 \\
         {[N\,{\sc ii}]$\lambda$6583/H$\beta$} & 4.72  & 4.96    & 5.10     & 5.26   & 5.33   & 5.23   & 4.99   \\
                                    & 3.54  & 3.71    & 3.90     & 3.91   &  3.84  & 3.94   & 3.75   \\
                                    &  $\pm$0.10/$\pm$0.29 & $\pm$0.12/$\pm$0.33 & $\pm$0.15/$\pm$0.41 & $\pm$0.17/$\pm$0.48 & $\pm$0.13/$\pm$0.36 & $\pm$0.17/$\pm$0.48 & $\pm$0.12/$\pm$0.34 \\
      {[S \,{\sc ii}]$\lambda$6716/H$\beta$} & 1.89  & 2.01    & 2.08     & 2.26   & 2.15   & 2.19   & 2.06   \\
                                    &  1.40 &  1.48   &  1.57    & 1.65   & 1.52   & 1.63   &  1.53   \\
                                    &  $\pm$0.05/$\pm$0.11 & $\pm$0.05/$\pm$0.13 & $\pm$0.07/$\pm$0.16 & $\pm$0.08/$\pm$0.20 & $\pm$0.06/$\pm$0.14 & $\pm$0.08/$\pm$0.20 & $\pm$0.06/$\pm$0.14 \\
       {[S\,{\sc ii}]$\lambda$6730/H$\beta$} &  1.40 &  1.50   & 1.57     & 1.66   & 1.60   & 1.62   & 1.55  \\
                                    &  1.04 &  1.11   & 1.19     & 1.21   &  1.13  & 1.20   & 1.15    \\
                                    &  $\pm$0.04/$\pm$0.08 & $\pm$0.05/$\pm$0.10 & $\pm$0.06/$\pm$0.12 & $\pm$0.07/$\pm$0.15 & $\pm$0.05/$\pm$0.10 & $\pm$0.07/$\pm$0.15 & $\pm$0.05/$\pm$0.10 \\    
         {HeII$\lambda$4686/H$\beta$} & 0.060 & 0.081  & 0.080  & 0.101 & 0.080 & 0.101 &  0.069  \\
                                      &  0.063 & 0.085  & 0.083  & 0.106 & 0.084 & 0.105 & 0.072  \\
                             & \small  $\pm$0.022/ & $\pm$0.024/ & $\pm$0.027/ & $\pm$0.031/ & $\pm$0.025/ & $\pm$0.031/ & $\pm$0.024/ \\ 
                             & \small  $\pm$0.034\hspace{0.15cm} & $\pm$0.033\hspace{0.15cm}  & $\pm$0.038\hspace{0.15cm}  & $\pm$0.037\hspace{0.15cm}  & $\pm$0.032\hspace{0.15cm}  & $\pm$0.033\hspace{0.15cm}  & $\pm$0.037\hspace{0.15cm}  \\  
\hline
\end{tabular}
\end{table*}

\section{Characteristic features of retired galaxies}
\label{results}

In this section, we present the results from the SED fitting of the stacked spectra using {\sc gandalf}, including the best-fit emission line spectra as templates for nebular emission of retired galaxies. We also present dust reddening of both the nebular and stellar emission.

\subsection{Nebular emission of the stacked spectra}
\label{results:nebular}

All the stacked spectra are fitted twice using {\sc gandalf}, first by adjusting the continuum of the stellar templates with  multiplicative polynomials and then by a reddening law (see Section~\ref{stacking:EMfitting}). In this section we provide the results for the former option as an emission line template for retired galaxies, but to test the robustness of the derived emission lines we compare them with the results for adopting a reddening law in the fitting procedure. First of all we compare the continuum correction for the two different cases. This is shown in Fig.~\ref{fig:contcorr} for the case of co-adding all galaxies selected with the PCA criterium (see Section~\ref{data}), where the continuum correction is presented as a function of wavelength for a 15 degree multiplicative polynomial (solid line) and the extinction curve from \citet{odonnell94} together with an measured E(B--V)=0.06 (dashed line). The overall shape of the two different corrections resemble each other very well, strengthening the argument that the polynomial correction is dominated by reddening effects as discussed in appendix~A in \citet{johansson14}. The main deviation occurs at the edges of the adopted wavelength range, possibly due to problems towards the limits of the observed wavelengths of both the SDSS and stellar template spectra. Such problems can be accounted for by adopting a high order multiplicative polynomial in order to measure weak emission lines as discussed in \citet{johansson14}, while strong emission lines are only mildly affected as discussed below.

Fig.~\ref{fit:HaLT0p4} (upper panel) presents the stacked spectrum, together with the best-fitting model adjusted with a multiplicative polynomial, from co-adding all galaxies selected with the PCA criterium (see Section~\ref{data}). This is a re-make of the upper panel in fig.~5 in \citet{johansson14}, but with the extended wavelength range for the new fits (see Section~\ref{stacking:EMfitting}). 
In the lower panel of Fig.~\ref{fit:HaLT0p4} the fit in three wavelength ranges containing prominent emission lines are high-lighted, namely; [O~{\sc ii}~]$\lambda$3727 ($3700-4100\AA$, lower left panel); H$\beta$, [O~{\sc iii}~]$\lambda$5007 ($4800-5200\AA$, lower middle panel); H$\alpha$, [N~{\sc ii}~]$\lambda$6583, [S~{\sc ii}~]$\lambda\lambda$6716,~6730 ($6400-6800\AA$, lower right panel).
As noted in \citet{johansson14}, the stacked spectrum is matched by the best-fit template to a very high precision, revealing a range of prominent emission lines. 
Appendix~A (available in the online version) contains corresponding figures for the rest of the stacked spectra (Fig.~A1-A6).

For each stacked spectrum we measure the strength of prominent nebular emission lines present in the 3700--6800 \AA\ wavelength window (see Section~\ref{stacking:EMfitting}). 
{Results are presented for lines that are unambiguously detected only (i.e. A/N $>6$, see Section~\ref{stacking:EMfitting}).}
Among the detected lines are the recombination Balmer lines H$\gamma$ {(A/N $=7.9-12.9$ for all stacked spectra)}, H$\beta$ {(A/N $=23.9-32.2$)} and H$\alpha$ {(A/N $=49.5-64.5$)}, and a range of forbidden lines including [O\,{\sc iii}]$\lambda$5007 {(A/N $=39.9-53.7$)} and [N\,{\sc ii}]$\lambda$6583 {(A/N $=69.3-88.4$, i.e. the strongest detected line)}.
The derived fluxes, both observed and corrected for extinction using the Balmer line decrement (see Section~\ref{stacking:EMfitting} and~\ref{results:dust}), are presented in Table~\ref{emisson} together with 1$\sigma$ errors. 

{Furthermore, to demonstrate the relative strengths of the detected lines, in Table~\ref{HbRatio} we list the line intensities relative to H$\beta$. 
While, as expected, no trend can be seen for the recombination Balmer lines, the general trend for the forbidden lines is that the relative strength to H$\beta$ increases slightly with stellar population age, both for the measured and reddening corrected ones. This increase varies between $\sim$10\% and $\sim$30\% for the different lines. Any further discussion of this trend requires a more detailed investigation which we defer for future work. 
For comparison with the results of \cite{johansson14}, at the end of Table~\ref{HbRatio} we include also the measured values of the He II 4686\AA/H$\beta$ ratio for each stack, although the He II 4686\AA\ line is not detected with sufficient confidence to satisfy the (quite conservative) requirement for being listed in Table \ref{emisson}. Note that increasing the wavelength range over which we fit our model spectra results in only slight ($< 1 \sigma$) discrepancies with our result in \cite{johansson14}.}

\begin{figure*}
\centering
\includegraphics[height=0.55\textheight,clip=true,trim=3cm 3cm 2cm 2cm, angle=180]{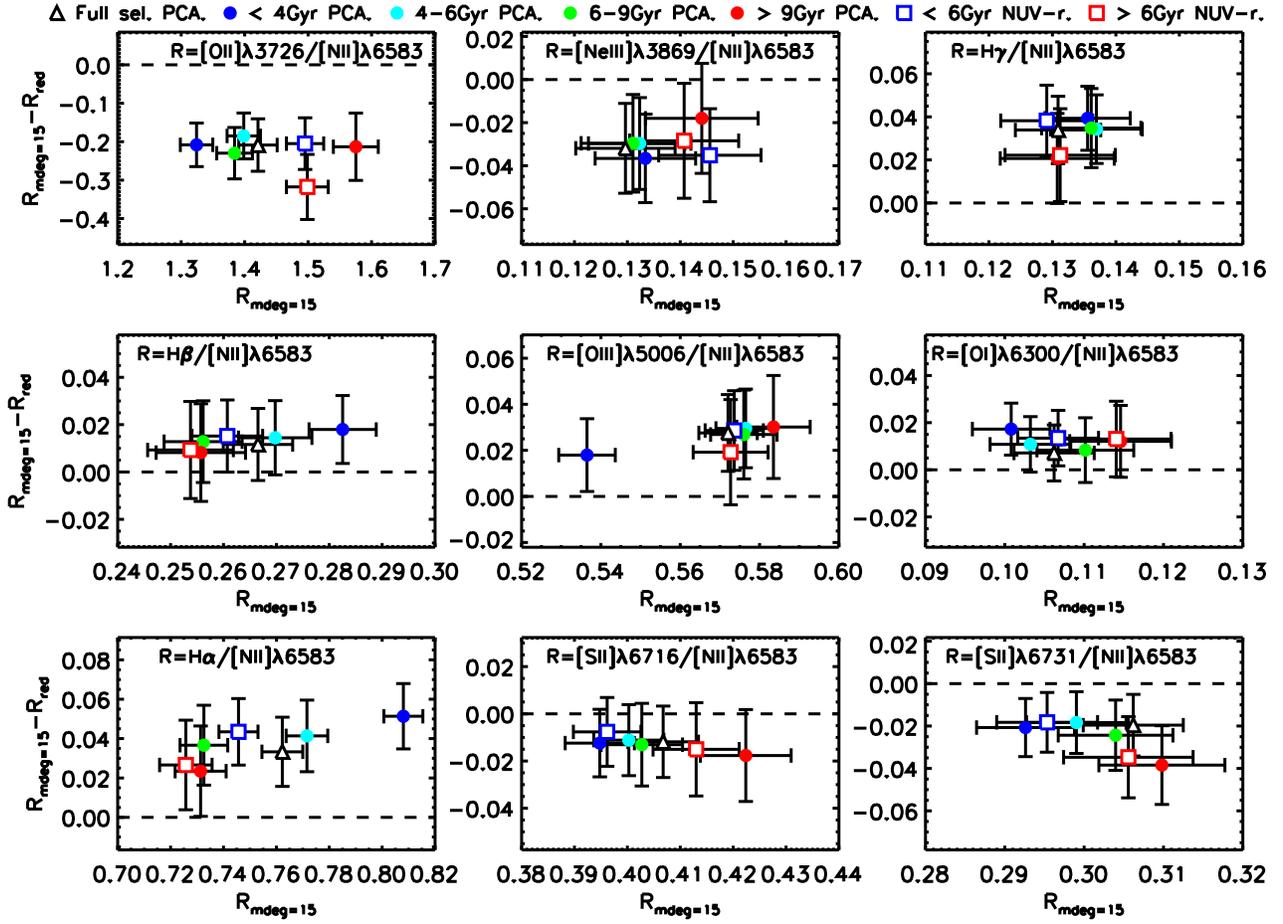}\\
\caption{Comparison of the difference in derived flux ratios using two different means of adjusting the derived stellar continuum: using a multiplicative 15-degree polynomial function ($\rm{R}_{15}$), and the reddening law of O'Donnell (1994) ($\rm{R}_{\rm{red}}$).}
\label{comp_flux}
\end{figure*}

\begin{figure}
\centering
\includegraphics[height=0.37\textheight,clip=true,trim=5cm 2.5cm 4cm 2.5cm]{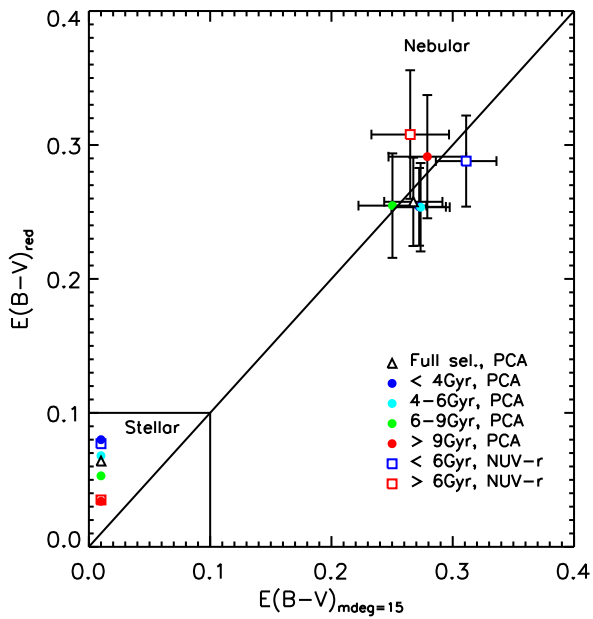}\\
\caption{E(B-V) for the different choices of running {\sc gandalf} (see Section~\ref{stacking:EMfitting}). Hence, E(B-V)$_{red}$ as found using the O'Donnell (1994) reddening law plotted against E(B-V)$_{mdeg=15}$ as found from the 15th degree multiplicative polynomial continuum correction, for the nebular (large area) and stellar spectra (small area) separately and for the differing age bins as indicated by the coloured data points with corresponding labels. Stellar E(B-V)$_{mdeg=15}$ values can not be derived when using the multiplicative polynomial continuum correction. Hence, these values are fixed to 0.01.}
\label{comp:ebmv}
\end{figure}
\begin{table*}
\center
\caption{Observed H$\alpha$/H$\beta$ ratios and color exceses for each of our summed spectra.}
\bigskip
\label{tab:reddening}
\begin{tabular}{lcccc}
\hline
\multicolumn{1}{c}{\bf Stack}  & \bf  E(B-V)$_\textrm{gas}$ & \bf H$\alpha$/H$\beta$ & \bf  E(B-V)$_\textrm{gas}$ & \bf  E(B-V)$_\textrm{stars}$ \\
                                                                            &         mdeg.=15$^a$                      &                        red.$^a$             &                red.$^a$                         &           red.$^a$  \\
\hline
$<$4 Gyr (PCA) &       0.272$\pm$0.022 & 3.74$\pm$0.11 & 0.25$\pm$0.029 & 0.080$\pm$0.004\\
4-6  Gyr (PCA) &      0.273$\pm$0.024 & 3.74$\pm$0.13 & 0.25$\pm$0.033 & 0.070$\pm$0.003\\
6-9  Gyr (PCA) &      0.250$\pm$0.028 & 3.74$\pm$0.15 & 0.26$\pm$0.039 & 0.060$\pm$0.004 \\
$>$9 Gyr (PCA) &       0.279$\pm$0.032 & 3.89$\pm$0.19 & 0.29$\pm$0.046 & 0.040$\pm$0.004 \\
$<$6  Gyr (NUV--r) &  0.311$\pm$0.025 & 3.87$\pm$0.14 & 0.29$\pm$0.034 & 0.080$\pm$0.004 \\ 
$>$6 Gyr (NUV--r) &    0.265$\pm$0.032 & 3.96$\pm$0.20 & 0.31$\pm$0.048 & 0.040$\pm$0.001 \\ 
Full sel.      &      0.267$\pm$0.024 & 3.75$\pm$0.13 & 0.26$\pm$0.033 & 0.060$\pm$0.001 \\ 
\hline
\end{tabular}
\flushleft
~~~~~~~~~~~~~~$^a$Method for adjusting the stellar continuum in {\sc gandlf/ppxf}, mdeg. = multiplicative degree and red. = reddening.
\end{table*}

To test the robustness of the derived fluxes, we compare the values derived for adjusting the stellar continuum with a multiplicative 15th order polynomial and using the reddening law of \citet{odonnell94} (see Section~\ref{stacking:EMfitting}). These comparisons are presented in Fig.~\ref{comp_flux} for the detected emission lines of all spectra as the ratio of all the lines to the strongest line in our spectra [N\,{\sc ii}]$\lambda$6583. The comparisons are shown as the difference between the two fitting options as a function of the ratio from fitting with a polynomial. We can see that the flux ratios typically differ within 3-$\sigma$ errors. The difference is slightly larger for [O\,{\sc ii}]$\lambda$3726 (upper left panel), which falls in the wavelength range where there is a more pronounced deviation in the continuum correction (see Fig.~\ref{fig:contcorr}).

\subsection{Dust reddening}
\label{results:dust}

In this section we investigate the measured dust extinctions affecting the nebular and stellar emission.  
{Table~\ref{tab:reddening} presents the color excess for the gas, E($B-V$)$_\textrm{gas}$, derived for the default set up of R$_v=3.1$, R$_\textrm{int}=2.86$ and for the observed H$\alpha$/H$\beta$ line ratios (see Section~\ref{stacking:EMfitting}). Table~\ref{tab:reddening} presents the results for both adopted options of adjusting the stellar continuum of the templates, i.e. using a multiplicative polynomial degree and reddening (see Section~\ref{stacking:EMfitting}). The H$\alpha$/H$\beta$ line ratios for the latter option were presented in Table~\ref{emisson}, while the corresponding numbers for the option of adjusting the stellar continuum using reddening are included in Table~\ref{tab:reddening}.}
The values for the two options agree {very well and are} within the 1$\sigma$ errors. {This is shown in the main panel of Fig~\ref{comp:ebmv}, where E(B-V)$_\textrm{gas}$ for the reddening case is plotted as a function of E(B-V)$_\textrm{gas}$ for the multiplicative polynomial case together with their corresponding 1$\sigma$ errors}. For our preferred set up of adjusting the continuum with a multiplicative degree of 15, the observed 
H$\alpha$/H$\beta$ ratio and E(B-V)$_\textrm{gas}$ values varies only mildly between the different stacks and have average values of 3.82 and 0.274 mags, respectively. The errors in the H$\alpha$/H$\beta$ ratio and E(B-V)$_{gas}$ are derived through Monte Carlo simulations using the errors on the individual fluxes.

When adjusting the continuum with a reddening law we do also derive the color excess E(B-V)$_\textrm{stars}$ for the extinction affecting the stellar emission, also presented in Table~\ref{tab:reddening}. Since the stacked spectra have very high S/N ratios, the stellar templates can be constrained to a very high precision through a number of strong absorption features in the wavelength window adopted for the fitting. Hence, the reddening affecting the stellar continuum can in turn be constrained to a high level. We find a significantly lower reddening factor affecting the stellar emission as compared to that affecting the nebular emission. The E(B-V)$_\textrm{stars}$ values fall in the range $0.04-0.08$ mag, with a trend of declining values with increasing stellar population age.
For the relation between the extinction affecting the nebular and stellar emission,  we find values in the range $0.14\textrm{E}(B-V)_{gas}<\textrm{E}(B-V)_{stars}<0.29\textrm{E}(B-V)_\textrm{gas}$ with an average of 
\begin{equation}
\textrm{E}(B-V)_{stars}=0.22\textrm{E}(B-V)_{gas}
\label{eq:ebmv_diff}
\end{equation}

\section{Discussion}
\label{disc}

\subsection{Emission-line diagnostics characterizing the physical conditions in the gas}
\label{disc:diffgass}

The [S~{\sc ii}~]$\lambda$6716/H$\alpha$ ratio for the different stacks is consistently $\sim$0.5. This confirms the assumption that the gas in the studied galaxies is in a diffuse ionized state, which is typically indicated by a [S~{\sc ii}~]$\lambda$6716/H$\alpha$ ratio $>0.2$ \citep{vo87}. 
{Assuming a gas temperature of $10^{4}K$, we find that for all age bins the [S II] $\lambda$6716/$\lambda$6731 ratio ($\approx$ 1.3) indicates an electron density of $\approx$ 120--150$\rm{cm}^{-3}$ \citep{McCall84,osterbrock89}. 
This is consistent within a factor of two with the density measured from the  observed ratio [O II] $\lambda$3729/$\lambda$3726 $\approx$1.1 \citep[see, e.g.,][]{Pradhan06}.
Furthermore, the estimated density values are broadly consistent with the results of previous studies of the warm neutral/warm ionized medium in retired galaxies \citep[see, e.g.][]{sarzi10,yan12}.
Discrepancies in density estimates between different line ratios are not important for the purpose of this paper.}

Beyond these simple diagnostics, further understanding of the warm ISM of retired galaxies requires a more precise understanding of the sources of ionizing radiation powering their emission-line regions. Unfortunately, there remains some doubt in this matter, although considerable progress has been made in demonstrating that post-AGB stars, and pAGBs alone, are primarily responsible for providing the needed ionizing background in most such galaxies \citep[e.g.][]{sarzi10, yan12, singh13, johansson14}, however other possibilities remain. In particular, \cite{woods} demonstrated that accreting, nuclear-burning white dwarfs would, in the canonical ``single-degenerate scenario'' for the production of type Ia supernovae (SNe Ia), necessarily provide the dominant contribution to the ionizing background supplied by old stars. Using the He II emission line diagnostic of \cite{woods}, \cite{johansson14} demonstrated that accreting white dwarfs cannot be so numerous as to either account for the total observed SN Ia rate in old stellar populations, nor dominate their ionizing background. In the following subsection, we expand on this result, using the similar (and more stringent) diagnostic of \cite{woods14}, using the [O I] $\lambda$6300 emission line. 

\subsection{Constraining the ionizing background in retired galaxies}
\label{disc:ionisation}

The nature of the ionizing sources powering the emission-line regions observed in early-type galaxies remains uncertain. Post-AGB stars are the favoured candidate ionizing source population on a variety of grounds \citep{binette94, stasinska08, sarzi10, yan12, Papaderos13, singh13}, however other ionizing sources should exist as well within the old, evolved stellar population. In particular, accreting, steadily nuclear-burning white dwarfs (WDs) are expected to number in the hundreds to thousands within a typical elliptical galaxy \citep[e.g.][]{Chen14}. With photospheric temperatures on the order of $\approx$1.5--10$\times 10^{5}K$ and spectra well-approximated by blackbodies \citep[in particular, without any sharp cutoff at the H I or He II ionization edges, see][]{Rauch10, woods}, accreting WDs could in principle constitute an important part of the ionizing background in early-type galaxies. 

A detailed calculation of the number and combined luminosity of accreting WDs within a given stellar population depends on as-yet uncertain details of binary stellar evolution. As a first approximation, we may ask what their luminosity would be if nuclear-burning WDs were the progenitors of type Ia supernovae (SNe Ia). In the so-called single-degenerate (SD) scenario, a carbon-oxygen (CO) WD accretes hydrogen-rich material from some main-sequence or red giant companion, and grows through steady nuclear-burning of this material at its surface until reaching the Chandrasekhar mass ($\rm{M}_{\rm{Ch}}$=1.4$M_{\odot}$), triggering an explosion as a SN Ia. Should this scenario hold for all SNe Ia in a given population, a certain amount of matter must be accreted by the population as a whole in order to bring the correct number of white dwarfs to $\rm{M}_{\rm{Ch}}$ per unit time, such that it matches the observed SN Ia rate \citep{gilfanov}. Since this matter must be processed through nuclear-burning at the surface, this gives us an estimate, in an average sense, of the total luminosity of the population as a whole

\begin{equation}
\rm{L}_{\rm{Ia}} = \epsilon \chi \Delta\rm{m}\dot N_{\rm{Ia}}(\rm{t}) \label{L_Ia_general}
\end{equation}

\noindent where $\epsilon \approx 6.4\times 10^{18}$erg/g is the energy release from nuclear-burning of hydrogen, $\chi = 0.72$ is the hydrogen fraction of solar-metallicity gas, $\Delta \rm{m}$ is the average mass accreted per SN Ia, and $\dot N_{\rm{Ia}}(\rm{t})$ is the observed SN Ia rate as a function of delay time (age of the starburst). The maximum ``birth'' mass of a CO WD is $\approx$ 1.1$M_{\odot}$ \citep{umeda99}, therefore $\Delta \rm{m}_{min}$ = 0.3$M_{\odot}$ is the minimum mass which must be accreted through nuclear-burning in order to reach $\rm{M}_{\rm{Ch}}$. However, since the WD mass function peaks closer to 0.6$M_{\odot}$, one may infer that the average mass accreted per WD should likely be far greater. 

Plugging in the appropriate values into eq. \ref{L_Ia_general}, we have

\begin{equation}
\rm{L}_{\rm{Ia}} = 3.5\times 10^{31}\left(\frac{\Delta\rm{m}}{0.3M_{\odot}}\right)\left(\frac{\rm{t}}{1\rm{Gyr}}\right)^{-1}\rm{erg/s}/\rm{M}_{\odot,\rm{formed}} \label{L_Ia_ES0}
\end{equation}

\noindent where we have used the recent delay-time distribution of \cite{mm12}, a $\rm{t}^{-1}$ fit to all the observed delay-time distribution compiled in their review. Note that the stellar mass referred to in the normalization of eq. \ref{L_Ia_ES0} is the total mass {formed}, which makes for a more convenient comparison with our photoionization calculations, as well as population synthesis calculations in general.  
In order to constrain the accreting, nuclear-burning WD population, we may then use this estimate of their luminosity, together with some reasonable assumption for their temperatures, in order to compare their combined spectra with that of pAGB stars. We can then compare predictions for optical emission line ratios with and without the inclusion of accreting WDs to the ionizing background.

\begin{figure}
\begin{center}
\includegraphics[height=0.23\textheight,clip=true,trim=2cm 2.5cm 2cm 2cm]{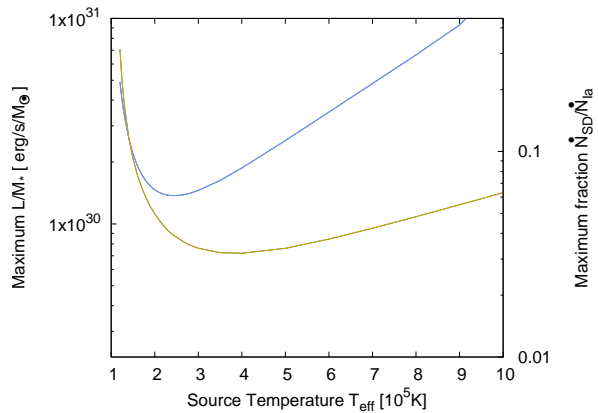}
\caption{Upper limits (90\% confidence) on the specific ionizing luminosity (left y-axis) and the (corresponding) possible contribution of accreting WDs to the SN Ia rate (right y-axis) as a function of the photospheric temperature. Yellow denotes our upper limit using the [O I]/H$\alpha$ diagnostic \citep{woods14}, while blue gives the upper limit from He II 4686/H$\beta$ \citep{woods}. For the contribution to the SN Ia rate, we assume a "minimal" average accreted mass per SN Ia of $\Delta$M = 0.3$\rm{M}_{\odot}$ (see text). Note that for each temperature, the upper limit is given excluding any emission at any other temperature, save for that from the ambient stellar population. \label{constraints}}
\end{center}
\end{figure}

To do so, we use the detailed photoionization code MAPPINGS III \citep[e.g.][]{Groves04}, together with the procedure outlined in \cite{woods}. In brief, we model the ionizing emission from the ``normally-evolving'' stellar population (principally originating from pAGBs) using the spectral synthesis models of \cite{bc03}. We assume a plane parallel geometry with an ionization parameter log(U) = -3.5 \citep[consistent with the observed O III/H$\beta$ ratio, see discussion in][]{johansson14}, with a warm gas phase density of n$\approx$100$\rm{cm}^{-3}$, and a gas phase metallicity of $\rm{Z}_{\rm{gas}}$ = $\rm{Z}_{\odot}$ (see previous subsection). {We assume that our model nebulae are ionization-bounded. Given the inferred volume densities and ionization parameter in their ISM, this is consistent with the observed H I column densities in retired galaxies.}. The observed SN Ia rate declines quickly with time from initial starburst (c.f. eq. \ref{L_Ia_ES0}), therefore to provide the strongest constraint on the progenitors of SNe Ia, we focus here on our youngest age bin, with a luminosity-weighted mean stellar age of $\approx$2.8 Gyr. In modelling the ionizing continuum from combinations of pAGB stars and SN Ia progenitor populations, \cite{woods14} \& \cite{johansson14} found that there was a negligible difference between using a single age stellar population (SSP) or a (more realistic) combination of old and young stellar populations with the same luminosity-weighted mean age. Therefore for simplicity we use a 2.8 Gyr-old SSP in modelling the ionizing continuum for our youngest age bin.

As shown in \cite{woods,woods14}, the most important lines for constraining the presence of an additional high temperature population are the He II 4686\AA\ recombination line, and the forbidden line [O I] 6300\AA. In our previous study \citep{johansson14}, we demonstrated that the observed He II 4686\AA/H$\beta$ ratio in our youngest stack already excludes accreting WDs with T $\approx 2$ -- $5 \times 10^{5}K$ from contributing significantly to the SN Ia rate. Here we extend those calculations to include our upper limits from the [O I] 6300\AA/H$\alpha$ ratio (see Fig. \ref{constraints}), as well as updating our He II constraint based on our new fit to the stellar continuum. 
{To this end we use the dependence of He II 4686\AA/H$\beta$ and [O I] 6300\AA/H$\alpha$ ratios upon the effective temperature  T$_{eff}$ predicted by the photoionization calculations of Woods \& Gilfanov (2013, fig.~7; 2014, fig.~2) for their default SD scenario model and combine them with the observed values of these ratios (0.063  and 0.125, see Tables~2 and~3).}
Note that our constraints are given for 90\% confidence upper limits on both the He II4686\AA/H$\beta$ and [O I] 6300\AA/H$\alpha$ ratios; the observed values of both are consistent with ionization by pAGB stars alone. 

As can be seen from Fig. \ref{constraints}, the observed [O I] 6300\AA/H$\alpha$ ratio provides a stronger or equivalent constraint than the He II 4686\AA/H$\beta$ ratio across the entire range in expected temperatures for accreting, nuclear-burning WDs. There are two reasons for this: first, the declining sensitivity of the He II diagnostic at high temperatures \citep[see][]{woods}, and second, [O I] 6300\AA\ is an intrisically stronger line than He II 4686\AA, as is H$\alpha$ stronger than H$\beta$. Therefore the [O I] 6300\AA/H$\alpha$ diagnostic is built on two more robustly detected lines. 

For both diagnostics, we have expressed our upper limit as a constraint on the total bolometric luminosity in excess of that provided by pAGB stars, as a function of blackbody temperature of the sources. Here we have normalized the ionizing luminosity per unit {instantaneous} stellar mass, where for a 2.8Gyr-old simple stellar population with Z=0.05, \cite{bc03} predict $\approx$0.554$M_{\odot}$ remaining in stars for every 1$M_{\odot}$ formed. Note that the upper limit at each temperature is to the exclusion of an additional excess at any other temperature. To convert this upper limit on the luminosity to an average accreted mass per WD, we may rearrange eq. \ref{L_Ia_ES0} together with the stellar mass (per solar mass formed) for a 2.8 Gyr-old stellar population from \cite{bc03} to find:

\begin{equation}
\Delta \rm{m}_{\rm{ul,2.8Gyr}} \approx 1.3\times 10^{-2} \left(\frac{\rm{L}_{\rm{Ia}}/\rm{M}_{*}}{10^{30}\rm{erg/s/\rm{M}_{\odot}}}\right) M_{\odot}
\end{equation}

\noindent This rules out accreting, nuclear-burning WDs as the progenitors of SNe Ia in old stellar populations, unless they account for only a fraction of the total. As a conservative estimate, we may set $\Delta \rm{m}_{\rm{Ia}}$ = 0.3$M_{\odot}$. In this case, our upper limits 
constrain the contribution of the accreting, nuclear-burning WDs to the total observed SN Ia rate to less than $\approx$6\% {for any source temperatures in the range $\rm{T}_{\rm{eff}} \approx (1.5 - 10)\times 10^{5}K$}. 

However, our ionizing luminosity constraints are not only applicable to SN Ia progenitors. Recently, population synthesis studies have predicted a steep evolution with time for accreting WDs regardless of the SN Ia rate \citep{Chen14}, and for low-mass X-ray binaries \citep{Fragos13}. In both cases, accreting compact objects are then expected to provide a strong contribution to the ionizing background in early-type galaxies of relatively young luminosity-weighted mean stellar age. Therefore, our results can inform our understanding of binary stellar evolution, providing significant constraints \citep[e.g.][]{Chen15}. 

\subsection{Color excess}
\label{disc:cover}

In Section~\ref{results:dust} we present derived values for the color excess for both the stellar and nebular emission of our stacked spectra. The color excess of the stellar continuum is $\sim$20 per cent of that of the nebular emission. 
This value is significantly lower than the ratio between E($B-V$)$_{stars}$ and E($B-V$)$_{gas}$ of 0.44 and 0.47 found for star forming galaxies in \citet{calzetti97b} and \citet{kreckel13}, respectively. 
In star forming galaxies mainly young O-stars are responsible for ionizing the nebular gas and these stars reside closer to the gas in HII-regions than older stars of the galaxies. 

\citet{kreckel13} study the nebular and stellar emission, similar to this work by using the fitting code {\sc gandalf}, of resolved nearby star-forming galaxies and find that the E($B-V$)$_s$/E($B-V$)$_g$ ratio depends on whether an HII-region (high star-formation rate indicated by strong H$\alpha$ emission) or regions of diffuse ionized gas (DIG), as determined by using the line ratio [S~{\sc ii}~]$\lambda$6716/H$\alpha$. In the HII-regions the E($B-V$)$_s$/E($B-V$)$_g$ ratio is $\sim$0.5, while it rises to $\sim$0.7 in the DIG-regions. This discrepancy is believed to arise due to a more uniform mixing of the dust in DIG-regions, while HII-regions do not dominate the stellar light in their vicinity. The latter is obviously dependent on the resolution of the observations. 

{On the contrary, our sample selection ensures that no young stars in HII-regions are present in the galaxies studied in this work, so that the dust should reside in a rather diffuse medium. Yet, the finding of a very low value for the E($B-V$)$_s$/E($B-V$)$_g$ ratio (around 0.2) suggests that the nebular and stellar emission in retired galaxies are not uniformly and similarly covered by dust, or otherwise a simple foreground dust screen morphology \citep[e.g.][]{calzetti97} would yield consistent E($B-V$)$_s$ and E($B-V$)$_g$ values.}

{Drawing from the typical distribution of ionised-gas and dust observed in nearby early-type galaxies \citep{sarzi06}, which generally trace each other and are distributed along lanes or disks, we can explain the observed discrepancy between the E($B-V$)$_s$ and E($B-V$)$_g$ values by assuming that only a fraction of the stellar regions encompassed by the 3''-wide apertures of the SDSS spectra are in fact covered by dust (e.g., by a dust lane or an inclined disk). Further assuming that along the line of sight towards these dusty regions the stellar and nebular flux are similarly attenuated and that elsewhere reddening by dust is essentially insignificant, as generally observed in early-type galaxies without dust lanes \citep[see, e.g., the reddening values for the "ordinary`` early-type galaxies of][]{Jeo2013}, then the range of E($B-V$)$_s$ values from 0.04 to 0.08 mags that were derived assuming a dust-screen morphology could in fact be due to between just 20\% to 35\% of encompassed stellar light suffering from the same dust extinction as observed for the ionised-gas (corresponding to E($B-V$)$_s=0.27$ mags).  Although derived under simple assumptions, such values for the fraction of light obscured by dust in early-type galaxies appears broadly consistent with the morphology of dust-lane ellipticals or of early-type galaxies with embedded gaseous disks \citep[see, e.g., the images in][]{goudfrooij94a,Mac96}.}

\section{Conclusions}
\label{conc}

We present emission line templates for retired galaxies of varying stellar age. While it is determined that a large fraction of early-type galaxies contain significant amounts of ionized gas, the ionizing sources present in such objects are still debated. Hence, high-quality observed emission line templates are important to further the investigation of ionizing sources in retired galaxies. 

The templates presented in this work are based on high signal-to-noise co-added spectra of $\sim11500$ SDSS galaxies, selected to have significant amounts of gas and to be devoid of star-formation and AGNs. The selected sample is divided into groups of varying stellar population age, estimated for the individual spectra using absorption line indices. The spectra in each group are co-added separately, resulting in stacked spectra with S/N$>800$. 
The fitting code {\sc gandalf} \citep{sarzi06} is adopted for measuring emission lines on the stacked spectra in the wavelength range $3700-6800$ \AA. 
 In this paper, the full range of unambiguously detected emission lines, i.e. lines with an A/N~$>6$, in the considered wavelength range are presented, including [O~{\sc ii}]~$\lambda$3726, [O~{\sc ii}]~$\lambda$3729, [Ne~{\sc iii}]~$\lambda$3869, H$\gamma$, H$\beta$, [O~{\sc iii}]~$\lambda$5007, [O~{\sc i}]~$\lambda$6300, H$\alpha$, [N~{\sc ii}]~$\lambda$6583, [S~{\sc ii}]~$\lambda$6716 and [S~{\sc ii}]~$\lambda$6731.

In \citet{johansson14}., the total flux in the He~{\sc ii}~$\lambda$4686 line was derived for the stacked spectra which are also investigated in this work. The strength of this line limited the populations of accreting white dwarfs 
to the level at which they can account for no more than $\sim5-10$ per cent of the observed SN~Ia rate through a comparison with models of photoionization for the temperature range $\sim(1.5-6)\times10^5$~K. This result was instead consistent with pAGBs being the sole ionizing source in all age bins.
In this paper, we extend this comparison for a wider temperature range, using the modeled [O~{\sc i}]~$\lambda$6300/H$\alpha$ line ratio from \citet{woods14}. The observed [O~{\sc i}]~$\lambda$6300/H$\alpha$ ratio constrains the luminosity of ionizing sources in retired galaxies to $\lesssim 6\%$ across the entire range in source temperatures ($\approx$1.5--10$\times 10^{10}K$) characteristic of accreting, nuclear-burning WDs. In future work, we will exploit a wider range of emission lines to further constrain the ionizing background in retired galaxies. 

Simultaneously with the fitting of the emission lines, the stellar continuum is also fitted using the MILES library of stellar spectra \citep{miles} as templates. The reddening of the stellar continuum and nebular emission can thus be estimated individually and compared to investigate the dust content of retired galaxies. We find that the extinction affecting the nebular emission is five times higher than that affecting the stellar continuum. This is in contradiction to the case of isotropically distributed dust and gas that would render similar extinction values for both the stellar and nebular emission.

\section*{ACKNOWLEDGMENTS}

J.J. would like to thank the Deutsche Forschungsgemeinschaft (DFG) for financial support. M.G. acknowledges hospitality of the Kazan Federal University (KFU) and support by the Russian Government Program of Competitive Growth of KFU. K.Oh acknowledges support from the National Research Foundation of Korea (SRC Program No. 2010-0027910) and the DRC Grant of Korea Research Council of Fundamental Science and Technology (FY 2012).

{}

\newpage

\appendix

\begin{figure*}

\section{Fits for all spectra}
\label{appendixA}

\centering
\includegraphics[height=0.4\textheight,clip=true,trim=0cm 2cm 0cm 2cm,angle=180]{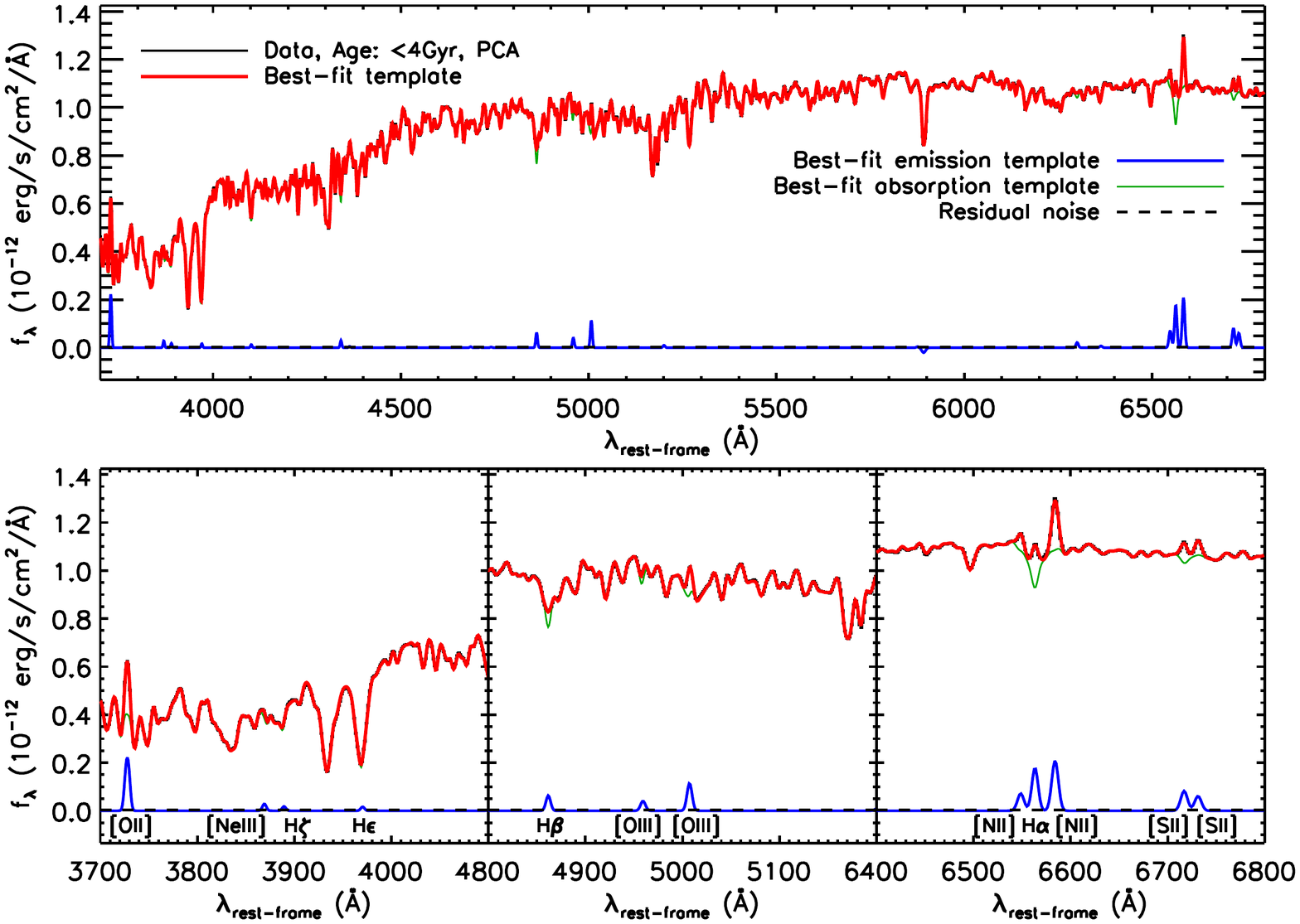}\\
\caption{Data and model of stacked spectrum with stellar ages $<$4 Gyr (PCA cut). Same as Fig.~\ref{fit:HaLT0p4} in the main manuscript, but
  for the spectrum obtained by co-adding the spectra of all galaxies
  in our sample with luminosity-weighted mean stellar ages $<$4 Gyr (2740
  objects).}
\label{fit:1}
\end{figure*}

\begin{figure*}
\centering
\includegraphics[height=0.4\textheight,clip=true,trim=0cm 2cm 0cm 2cm,angle=180]{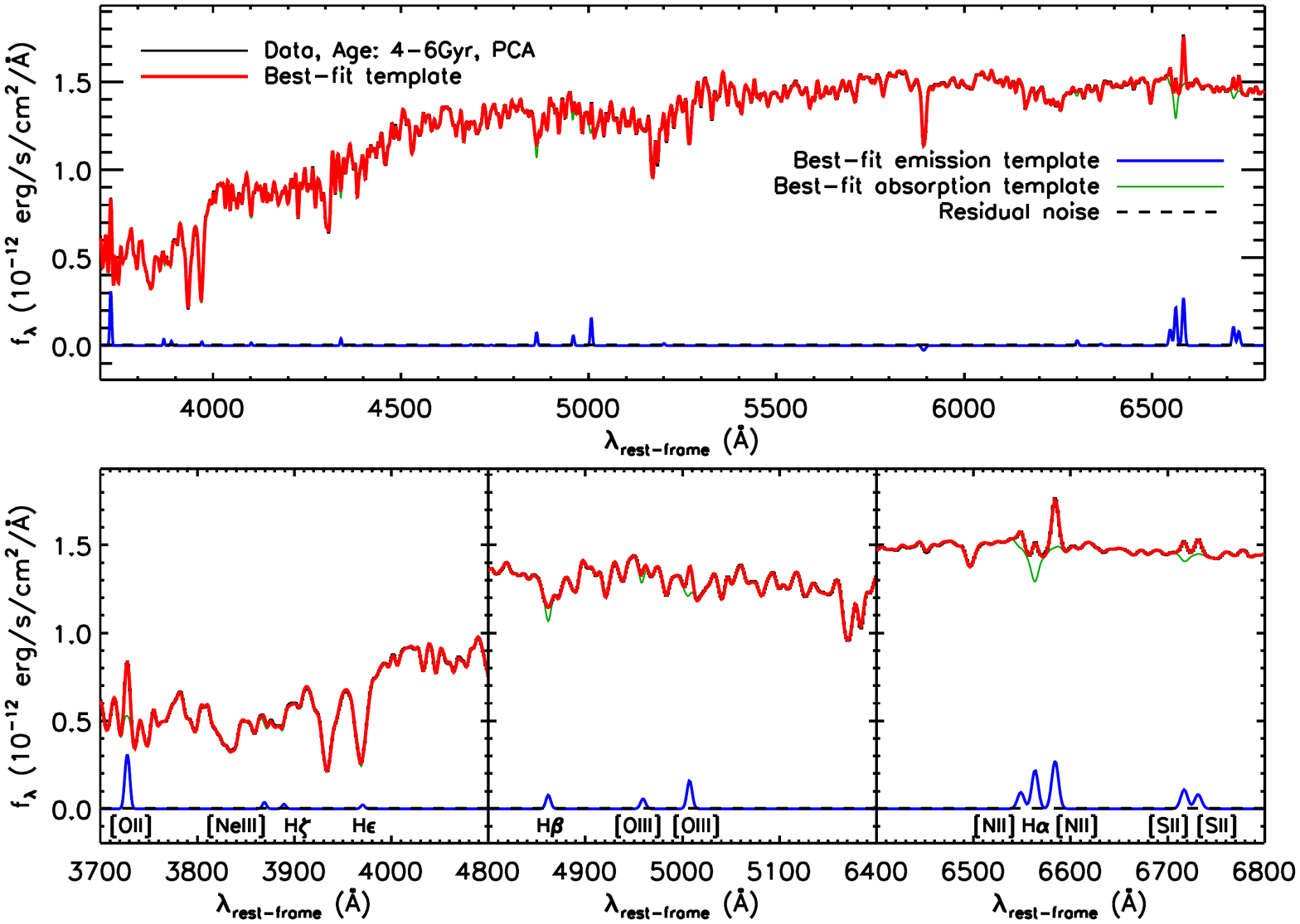}\\
\caption{Data and model of stacked spectrum with stellar ages between 4 and 6 Gyr (PCA cut). Same as as Fig.~\ref{fit:HaLT0p4} in the main manuscript and
  Fig.~\ref{fit:1}, but for the spectrum obtained by co-adding the
  spectra of all galaxies in our sample with luminosity-weighted mean
  stellar ages between 4 and 6 Gyr (3380 objects).}
\label{fit:2}
\end{figure*}

\begin{figure*}
\centering
\includegraphics[height=0.4\textheight,clip=true,trim=0cm 2cm 0cm 2cm,angle=180]{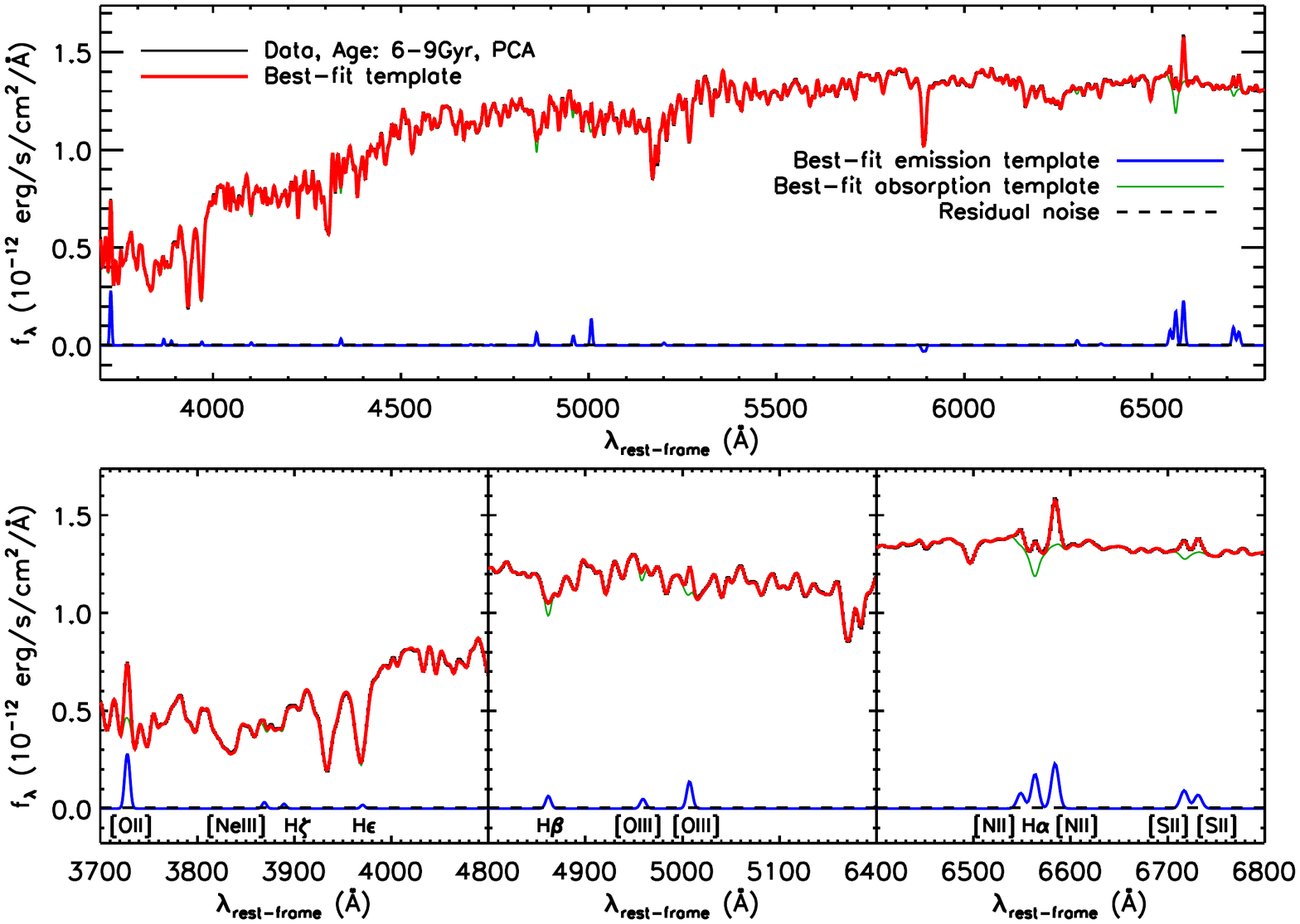}\\
\caption{Data and model of stacked spectrum with stellar ages between 6 and 9 Gyr (PCA cut). Same as Fig.~\ref{fit:HaLT0p4} in the main manuscript and
  Figs.~\ref{fit:1}-\ref{fit:2}, but for the spectrum obtained by
  co-adding the spectra of all galaxies in our sample with
  luminosity-weighted mean stellar ages between 6 and 9 Gyr
  (2729 objects).}
\label{fit:3}
\end{figure*}

\begin{figure*}
\centering
\includegraphics[height=0.4\textheight,clip=true,trim=0cm 2cm 0cm 2cm,angle=180]{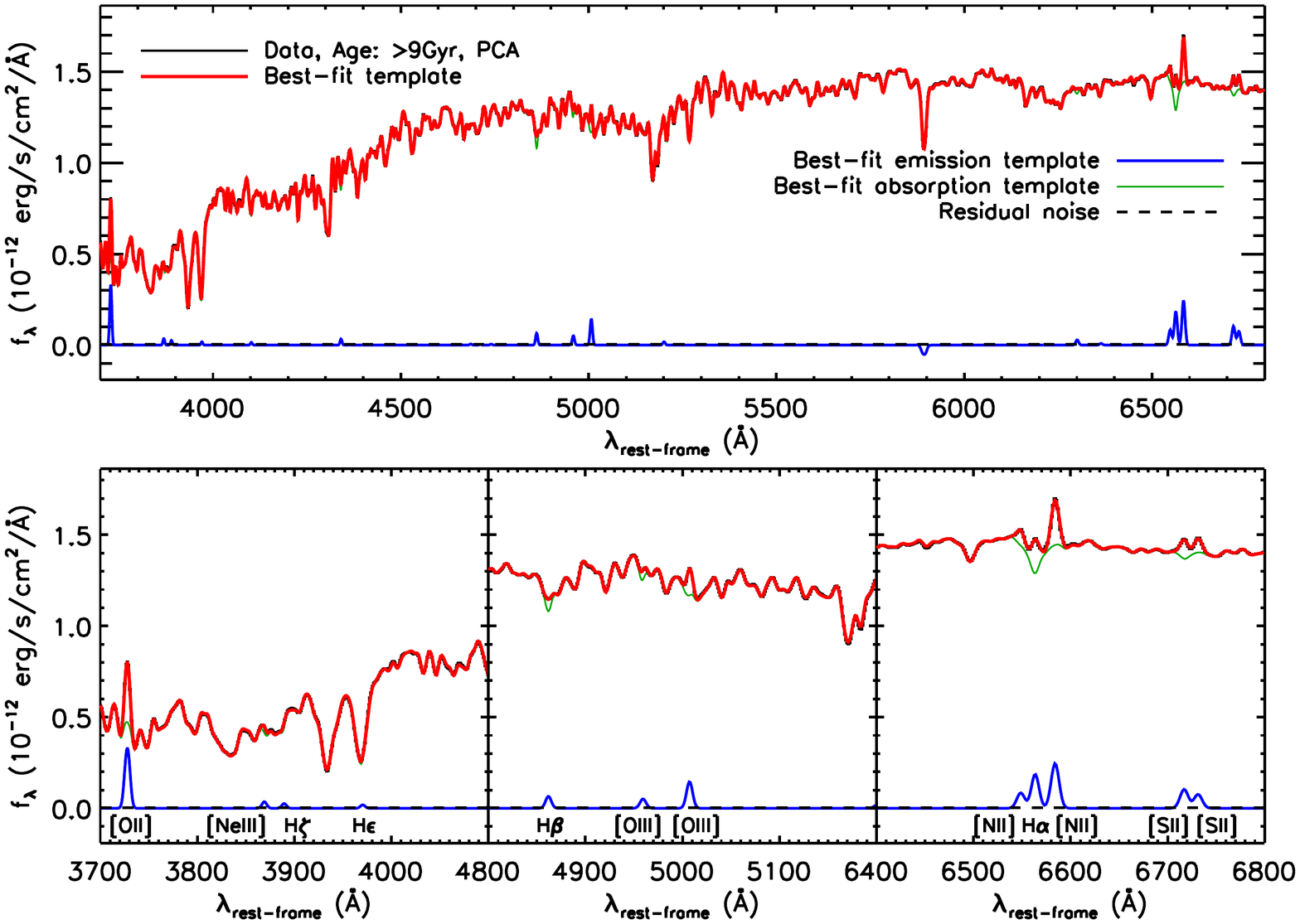}\\
\caption{Data and model of stacked spectrum with stellar ages $>$9 Gyr (PCA cut). Same as Fig.~\ref{fit:HaLT0p4} in the main manuscript and
  Figs.~\ref{fit:1}-\ref{fit:3}, but for the spectrum obtained by
  co-adding the spectra of all galaxies in our sample with
  luminosity-weighted mean stellar ages $>$9 Gyr (2744 objects).}
\label{fit:4}
\end{figure*}

\begin{figure*}
\centering
\includegraphics[height=0.4\textheight,clip=true,trim=0cm 2cm 0cm 2cm,angle=180]{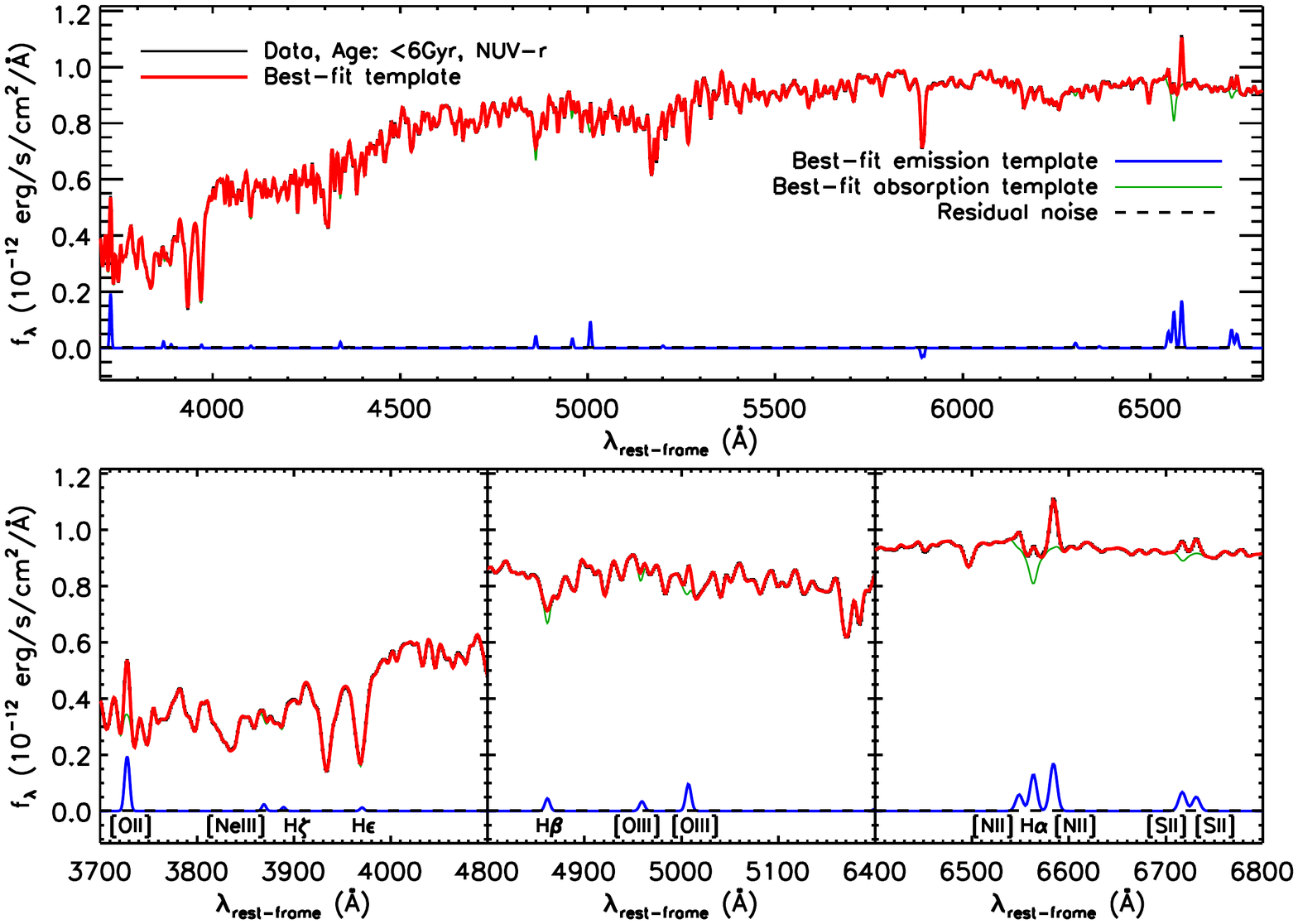}\\
\caption{Data and model of stacked spectrum with stellar ages $<$6 Gyr (NUV-r cut). Same as Fig.~\ref{fit:HaLT0p4} in the main manuscript and
  Figs.~\ref{fit:1}-\ref{fit:4}, but for the spectrum obtained by
  co-adding the spectra of all galaxies in our sample with
  luminosity-weighted mean stellar ages $<$6 Gyr (1953 objects).}
\label{fit:NUVr1}
\end{figure*}

\begin{figure*}
\centering
\includegraphics[height=0.4\textheight,clip=true,trim=0cm 2cm 0cm 2cm,angle=180]{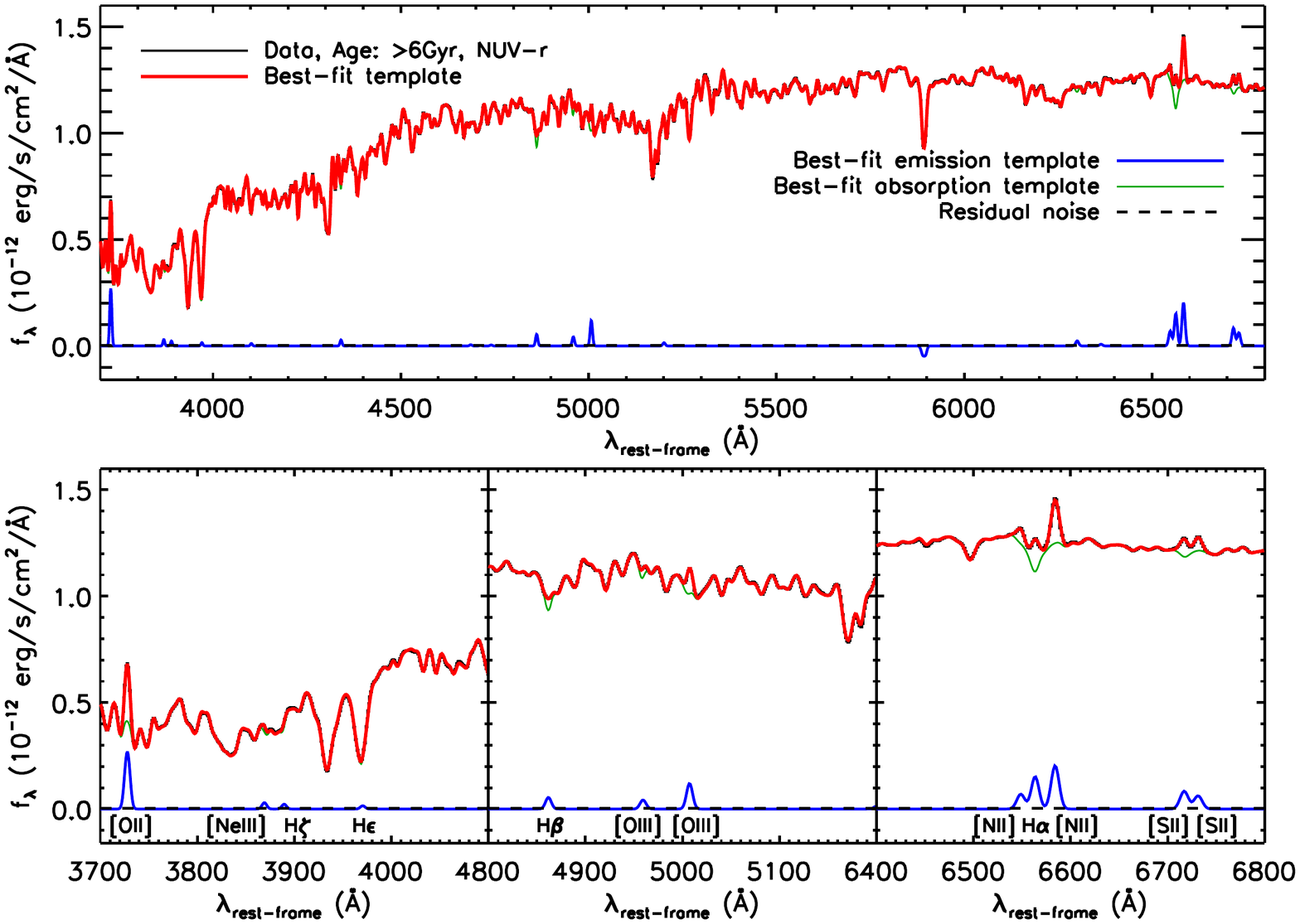}\\
\caption{Data and model of stacked spectrum with stellar ages $<$6 Gyr (NUV-r cut). Same as Fig.~\ref{fit:HaLT0p4} in the main manuscript and
  Figs.~\ref{fit:1}-\ref{fit:NUVr1}, but for the spectrum obtained by
  co-adding the spectra of all galaxies in our sample with
  luminosity-weighted mean stellar ages $>$6 Gyr (2108 objects).}
\label{fit:NUVr2}
\end{figure*}

\label{lastpage}

\end{document}